\documentclass[11pt,a4paper]{article}
\usepackage{graphicx} 
\usepackage[dvipsnames]{xcolor}
\usepackage[margin=2.6cm,nohead]{geometry}
\usepackage{amssymb}
\usepackage{amsmath}
\usepackage{booktabs}
\usepackage{cite}
\usepackage[hidelinks, colorlinks=true, allcolors=blue]{hyperref}
\usepackage{cleveref}
\usepackage{authblk}
\usepackage{algorithm}
\usepackage[noend]{algorithmic}

\usepackage{enumitem} 
\usepackage[export]{adjustbox} 
\usepackage{xspace}

\renewcommand{\d}{\ensuremath{\mathrm{d}}}

\newcommand{\Sunshine}{\textsc{Sunshine}\xspace}

\newcommand{\Vincia}{\textsc{Vincia}\xspace}
\newcommand{\EventTwo}{\textsc{Event2}\xspace}

\newcommand{\FigRef}[1]{Fig.~\ref{#1}\xspace}
\newcommand{\figRef}[1]{fig.~\ref{#1}\xspace}
\newcommand{\figsRef}[1]{figs.~\ref{#1}\xspace}
\newcommand{\eqRef}[1]{eq.~\eqref{#1}\xspace}
\newcommand{\eqsRef}[1]{eqs.~\eqref{#1}\xspace}
\newcommand{\secRef}[1]{sec.~\ref{#1}\xspace}
\newcommand{\secsRef}[1]{secs.~\ref{#1}\xspace}
\newcommand{\SecRef}[1]{Sec.~\ref{#1}\xspace}

\title{\bf Sudakov evolution without unitarity}
\author[1]{J.~Altmann}
\author[2]{H.~T.~Li}
\author[1]{L.~Scyboz}
\author[1]{P.~Skands}
\affil[1]{\normalsize School of Physics \& Astronomy, Monash University, Clayton VIC-3800, Australia}
\affil[2]{\normalsize School of Physics, Shandong University, Jinan, Shandong 250100, China}
\date{\today}

\begin{document}
\maketitle

\begin{abstract}
We present a method for sampling singular functions defined on (nested) multiparticle phase spaces, based on a generalisation of parton-shower phase-space generation techniques. At the heart of the method are three key ingredients: 1) the Sudakov sampling by which shower-style calculations sweep across phase space in an ordered manner, from hard to soft; 2) the sequential nesting of multiparticle phase spaces; and 3) the factorisations obeyed by singular multiparton amplitudes on the edges of these phase spaces. We demonstrate a C++ implementation of the proposed algorithm, dubbed \Sunshine,\footnote{Loosely derived from \textbf{Sudakov Nesting of Hard Integrals.}} for hadronic Z decays, and use it to test the tree-level accuracy of the \textsc{Vincia} sector shower through ${\cal O}(\alpha_s^2)$. \\[0.5cm]
\begin{center}
\framebox{\includegraphics*[height=0.11\textheight]
{sunshine\_2.jpeg}}\\
\end{center}
\end{abstract}

\setcounter{tocdepth}{2}
\tableofcontents
\clearpage

\section{Introduction}
\label{sec:intro}

This work originated in the study of tree-level expansions of parton showers, which are among the main diagnostics tools to assess the accuracy of parton showers at higher orders~\cite{Nagy:2009re,Skands:2009tb,Giele:2011cb,Dasgupta:2018nvj,Forshaw:2020wrq,Preuss:2024vyu,Hoche:2024dee,vanBeekveld:2024wws}. So far, fixed-order expansions of showers have typically been examined \emph{not} by directly running the actual shower algorithm, but instead by constructing explicit products of (sums over) their radiation kernels and phase-space mappings. This should of course be equivalent but does require writing additional code, which can be complex, and also introduces the risk that the actual shower differs in some unanticipated way from the (separately constructed) direct expansion. In this paper, we propose a simple and general technique for obtaining fixed-order expansions straight from shower generators.

The crucial observation is that the Sudakov factors (or, really, no-emission probabilities) produced by parton-shower algorithms express the principle of detailed balance: for each $(n+1)$-parton configuration created by a shower, a corresponding $n$-parton event is ``annihilated''. This ensures that the shower operator is \emph{unitary}: the action of the shower does not change the total (inclusive) cross section. Thus, in general, the ``weight'' of an $(n+m)$-parton configuration produced by a shower that starts from an $n$-parton configuration is, schematically:
\begin{equation}
w_{n+m} ~=~w_n\prod_{i=n}^{n+m-1}{\cal P}_{i\mapsto i+1}\,\Delta(t_{i},t_{i+1})\,,
\end{equation}
where $w_n\propto |M_n|^2$ is the weight of the starting ($n$-parton) configuration, ${\cal P}$ represents (fixed-order) branching kernels of the shower at hand (e.g., DGLAP splitting kernels~\cite{Gribov:1972ri,Altarelli:1977zs,Dokshitzer:1977sg} or dipole/an\-tenna functions~\cite{Gustafson:1987rq,Catani:1996jh,Kosower:1997zr,Gehrmann-DeRidder:2005btv}), and the Sudakov factors $\Delta \sim \exp(-\int {\cal P})$ represent no-emission probabilities between the sequential (nested) branching scales, $t_{i+1} < t_{i}$, with the shower evolution variable $t$ a measure of quantum mechanical resolution scale analogous to a jet-resolution scale. (If a shower algorithm has multiple ways to produce the given $(n+m)$-parton configuration, one must further sum over all such paths.)

In the context of studying \emph{tree-level expansions} of parton-shower algorithms, one would ideally like to ``remove'' the Sudakov factors, and study just the tree-level expansion, 
\begin{equation}
w_{n+m}^{(0)} ~=~w_n^{(0)}\prod_{i=n}^{n+m-1}{\cal P}^{(0)}_{i\mapsto i+1}\,,
\label{eq:schemTree}\end{equation}
where the superscript $(0)$ indicates expansion to the first non-trivial order. This is what has traditionally been done by direct constructions of the product $\prod{\cal P}_{i}$ in \eqRef{eq:schemTree}. In this work, we point out that this can also be accomplished by running the shower algorithm in a dedicated mode in which the principle of detailed balance is not imposed. 

We present an explicit implementation of this proposal in the framework of \textsc{Pythia}~8.3~\cite{Bierlich:2022pfr}, with detailed validations, and use it to study tree-level expansions of the \textsc{Vincia} sector shower~\cite{Brooks:2020upa}, for hadronic $Z$ decays with and without tree-level matrix-element corrections up to second order. We note that most of our discussion here will focus on QCD but the general principle also applies to QED showers, or indeed to any Sudakov-based formalism that implements detailed balance. 

In \secRef{sec:sampling-allorders}, we summarise the standard method of Sudakov-based sampling of nested phase spaces in shower-style algorithms, and establish some notation that we shall use throughout. In \secRef{sec:SudFO}, we discuss the problem of obtaining fixed-order expansions from such algorithms in detail, present the \textsc{Sunshine} algorithm, and give a mathematical proof that it produces the desired tree-level-expanded weights. We also discuss subtleties associated with ensuring full phase-space coverage, shower algorithms with general $n\mapsto n+m$ branchings (for $m\ge 2$), the infrared cutoff in shower algorithms, and finally present a set of validation checks which demonstrate that our implementation of the proposed algorithm is self-consistent. \SecRef{sec:treeShower} illustrates the application of the algorithm to tree-level tests of the \textsc{Vincia} sector shower at 1$^\mathrm{st}$ and 2$^\mathrm{nd}$ order, comparing it against both analytic results~\cite{Ellis:1980wv,Gehrmann-DeRidder:2005btv} and to results obtained with \textsc{Event2}~\cite{Catani:1996jh,Catani:1996vz}. Finally, in \secRef{sec:conc}, we conclude and give an outlook. The algorithm as we implemented it in \textsc{Pythia} is given in appendix~\ref{app:sunshine-algo}.

\section{Standard Sudakov sampling of nested phase spaces}
\label{sec:sampling-allorders}

Mathematically, parton-shower algorithms are formulated as Markov chains which operate on partonic states. These states consist of a set of partons specified by their momenta and information about their flavours and other internal quantum numbers. The latter typically includes colour flow in a leading-colour (LC) limit.\footnote{Conventionally represented by so-called Les Houches colour tags~\cite{Boos:2001cv,Alwall:2006yp}.} Optionally, helicity information may also be present and may then be used to generate helicity-dependent showers as in refs.~\cite{Larkoski:2013yi,Fischer:2017htu}. (More sophisticated treatments of spin-correlation and polarisation effects are beyond the scope of our discussion here.) 

For each input state, one seeks to generate radiation patterns that have eikonal factors as their soft limits and DGLAP kernels as their collinear limits (which we refer to collectively as infrared limits),
\begin{equation}
     \mathrm{ant}_{n\to n+m} \sim \frac{|M_{n+m}|^2}{|M_n|^2}\,,
\end{equation}
where $\mathrm{ant}_{n\to n+m}$ expresses an approximation which, at the very least, should  reproduce the leading pole structure(s) of the matrix elements. For definiteness, here we focus on the sector-antenna approach~\cite{Kosower:1997zr,Kosower:2003bh,Larkoski:2009ah,Lopez-Villarejo:2011pwr,Skands:2020lkd,Brooks:2020upa}. 

Starting from a Born state which we denote $\Phi_0$,
the veto algorithm (see, e.g. refs.~\cite{Mrenna:2016sih,Kleiss:2016esx,Bierlich:2022pfr}) is used to generate (unweighted) branchings distributed according to
\begin{equation}
 \frac{\d {\cal P}}{\d \Phi_{1}} = \mathrm{ant}_{0\mapsto 1}\,\Delta_0(t_0, t_1)\,|M_0|^2\,,
\end{equation}
where $M_0(\Phi_0)$ is the Born-level matrix element,
$t$ represents the evolution variable of the shower with   $t_0(\Phi_0)$ the starting scale and $t_1(\Phi_1) < t_0$ 
the evolution scale associated with the $0\mapsto 1$ branching, and $\Delta_0(t_0,t_1)$ is the Sudakov factor, that expresses the probability that the shower does not generate any branchings in the interval $t\in [t_0, t_1]$.

In a shower Markov chain this sampling is iterated, to build up higher multiplicities with a compound probability distribution proportional to 
\begin{equation}
\frac{\d {\cal P}}{\d \Phi_{n}} = |M_0|^2 \, \prod_{i=0}^{n-1} \mathrm{ant}_{i\to i+1}\,\Delta_i(t_i, t_{i+1})\,.
\label{eq:ps}
\end{equation}

The product of Sudakov factors explicitly represents the resummation carried out by the shower between each of the successive scales $t_i$. In this work, we are interested in modifying this sampling to eliminate the Sudakov factors, thus making the resulting sampling proportional to the tree-level expansion of the shower instead. 

We note that a variant of the above that will be especially interesting to us is one in which iterated matrix-element corrections (MECs)~\cite{Giele:2011cb} are applied during the shower evolution. This corrects the products of antenna functions in the above to ratios of matrix elements, such that
\begin{equation}
\frac{\d {\cal P}}{\d \Phi_{n}} = |M_{n}|^2 \, \prod_{i=0}^{n-1} \Delta^\mathrm{MEC}_i(t_i, t_{i+1})\,,
\end{equation}
where the superscript on the Sudakov factor emphasises that the effect of MECs is exponentiated and hence also modifies the probability densities in the Sudakov factors.

\section{Sudakov-based tree-level sampling}\label{sec:SudFO}

In the context of fixed-order applications, the standard Sudakov sampling method presented above suffers from four main drawbacks: 
\begin{enumerate}
\item Each generated phase-space point is weighted not only by the desired (product of nested) singular function(s) but also by all-orders Sudakov factors. In a parton-shower framework, these are of course desired and represent the resummations carried out by the shower, but for fixed-order applications they need to be expanded, either manually, or automatically as we propose in this section.
\item While a conventional Sudakov-style sampling could be highly efficient for intermediate-scale branchings, the strong Sudakov suppression of the singular regions themselves would translate to a poor sampling of those regions, undesired in a fixed-order context.
\item Conventional shower algorithms only account for iterated $n \to n+1$ branchings which, moreover, are strongly ordered in the shower evolution variable. In general, this implies that some (ordinarily small) phase-space regions are not covered. These so-called shower dead zones typically represent regions in which the shower splitting kernels and/or Sudakov factors do not produce good approximations to the structure of corresponding fixed-order amplitudes. Thus, from a pure-shower point of view, zero may be consistent with the best approximation the shower can offer there. Nevertheless, for matrix-element corrections and fixed-order applications, one would prefer a prescription that covers all of phase space. Failing that, any regions not reached by the Sudakov-based sampling would need to be populated by an alternative generator. This would both complicate the approach technically and presumably somewhat obviate any efficiency gains to be had from the Sudakov sampler. 
\item Conventional shower algorithms are terminated at an arbitrary but explicit IR cutoff, the lowest reasonable value for which is normally taken to be the hadronisation scale, in the range $\Lambda_\mathrm{QCD}$ to 1 GeV. In shower Monte Carlo applications, this is not a problem since hadronisation takes over below that scale. In a fixed-order context such a cutoff can be thought of as a slicing scale, which requires careful treatment. 
\end{enumerate}

In terms of a fixed-order application, the first two issues are essentially undesired side effects of the Sudakov-based sampling. In sec.~\ref{sec:newAlgo}, we propose an adaptation of the Sudakov algorithm which addresses these issues, by removing the constraint of unitarity from the evolution. 
In sec.~\ref{sec:fullPS}, we present two ways to address the third issue, to ensure full phase-space coverage: 1) smooth ordering~\cite{Giele:2011cb,Hartgring:2013jma}, and 2) direct $n\to n+m$ branchings~\cite{Li:2016yez}.

At the technical level, the fourth issue is mainly caused by shower resummations using a running $\alpha_s(p_\perp^2)$ to realise the best possible logarithmic accuracy~\cite{Catani:1990rr}. This makes it impossible to ``run across'' the Landau pole (unless one employs an IR-regulated coupling). As mentioned above, this is normally regarded as a feature, not a defect, since below the hadronisation scale a purely perturbative framework is anyway doomed to fail. In our context here, we will be using an $\alpha_s(\mu^2)$ with $\mu$ of the order of the scale of the hard process, which by definition will never conflict with $\Lambda_\mathrm{QCD}$ and will always be independent of any unresolved branchings. Just like in standard fixed-order applications, this will allow us to operate with a formal cutoff that is  limited only by machine precision. This is demonstrated in \secRef{sec:sudakov-tests}. In \secsRef{sec:speed} and \ref{sec:stability} we comment on speed and numerical stability aspects respectively. 

\subsection{Sudakov evolution without unitarity}
\label{sec:newAlgo}

In conventional Markov chains such as the shower evolution algorithm outlined above, the \emph{properties} of the state being evolved do, in general, change during the evolution but the \emph{weight} of that state (e.g., in terms of the fraction it represents of a total cross section) does not. This is sometimes referred to as detailed balance and reflects the fact that the evolution is unitary, i.e., conserves probability. For example, attaching a shower to a Born-level parton configuration will typically increase the multiplicity of the final state, as the shower generates branchings, but ultimately each input configuration produces one output configuration, with the ``output event'' having  the same weight as the input one. 

As a consequence, the summed weight of states that remain unchanged diminishes as the evolution proceeds. Detailed balance means that each lower-multiplicity state that generates a higher-multiplicity one is ``annihilated'' as the higher-multiplicity state is ``created''. This is nothing but a rewording of the familiar exponential decay law in nuclear physics (only nuclei that have not already decayed are available to decay at each instant in time), or closer to the problem at hand, the Sudakov factor on which shower algorithms are based.

Thus, to remove the Sudakov factors entirely, and thereby also eliminate the effects of resummation and its associated all-order weights, one simply has to never annihilate a state.\footnote{We note that a similar principle is used in the generation of the event-by-event normalisation factor, $n_B$, in the simplest version of the ESME approach to NLO matching~\cite{vanBeekveld:2025lpz}, and in merging methods at NLO such as NL$^3$ and extensions thereof~\cite{Lavesson:2008ah,Lonnblad:2012ix}.}
To adapt a Sudakov-style evolution algorithm to fixed-order sampling, we therefore propose to maximally violate the principle of detailed balance. That is, each time a new, higher-multiplicity state is created, one does \emph{not} annihilate the previous, lower-multiplicity one. In this way, each input event can in principle turn into arbitrarily many output events, each of which has the same weight as the input one had. This is the main technical difference with respect to conventional Sudakov-style evolution, which we elaborate on below. 

We also point out that if the Born-level input sample is unweighted, i.e., with all events having the same weight, then this will also be true of the output sample produced by this algorithm. The \emph{number} of events in the output sample will be larger than in the input sample, but the \emph{weight} of each event will be the same. One can of course also formulate various weighted versions of the algorithm, but since unweighting can be a difficult task in its own right we consider it a nice feature that the baseline algorithm will produce an unweighted output sample if given an unweighted Born-level sample as input. 

The basic setup is the same as in the previous section; we assume that the Born-level phase-space generation is not the main bottleneck one wants to address, i.e., that one already has a well-optimised Born-level phase-space generator. We also assume this Born-level generator is capable of assigning, or can fairly easily be supplemented by LC colour flows that can be used as the starting point for an LC antenna coverage of the subsequent radiative phase spaces.
One then commences the event evolution from the phase-space maximum $t_0$, with the following rules:
\begin{itemize}
\item When the shower produces a branching, the event evolution bifurcates: instead of replacing the old (pre-branching) event by the new (post-branching) one, both events are retained, each keeping the same weight as the original one had (thus doubling the total weight equivalent of the corresponding input event). Technically, in our implementation, we save the post-branching event to a buffer (along with the evolution scale $t$ at which it was created). We then veto the branching and continue the shower evolution of the pre-branching event from the scale $t$ downwards. 
\item Whenever the shower evolution of an event terminates (by reaching either the IR cutoff or a user-specified maximum parton multiplicity), the next event in the above-mentioned buffer is processed, starting from the bifurcation scale. (Note that this induces correlations between successive events; if this is not desired then the buffer should be accumulated during a full run and then shuffled before processing.)
\end{itemize}
The technical algorithm is elaborated upon in appendix~\ref{app:sunshine-algo}.
The procedure given above has the following two implications:
\begin{itemize}
\item Each Born-level input event produces a ``tree'' of output events, via bifurcations. Since the event weights are not modified, the weight of the original Born event is effectively multiplied by a number equal to the number of bifurcations produced. 
\item Since the unbranched events are always kept, the Sudakov factors that express the survival probability in the ordinary shower are absent: the generated probability distribution is proportional to the tree-level shower expansion,
\begin{equation}
\frac{\d {\cal P}}{\d \Phi_{n}} = |M_0|^2 \, \prod_{i=0}^{n-1}\mathrm{ant}_{i\mapsto i+1}\,.\label{eq:tree}
\end{equation}
\end{itemize}
We shall now give a proof of eq.~(\ref{eq:tree}). Each choice we can make at each bifurcation point leads to a distinct event sample. In each such sample, every event will have the same weight as its Born-level starting configuration had. The sample produced by following all of the branchings as one would normally do in a shower produces the same distribution as in eq.~(\ref{eq:ps}),
\begin{equation}
\frac{\d {\cal P}_{00\ldots 0}}{\d \Phi_n} = |M_0|^2 \prod_{i=0}^{n-1} \mathrm{ant}_{i\mapsto i+1}\Delta_i(t_i, t_{i+1})~\,, \label{eq:dP00}
\end{equation}
where the list of subscripts on $\cal P$ denotes the number of vetoed branchings we consider at each $i$. 
If we instead follow the branch that resulted from vetoing, say, branchings from the Born-level event $m$ times before following all subsequent branchings without vetoing, we get samples with weights 
\begin{equation}
\frac{\d {\cal P}_{m0\ldots 0}}{\d \Phi_n} ~=~ \frac{\d {\cal P}_{00\ldots 0}}{\d \Phi_n} 
\prod_{j=1}^m\int^{\tilde{t}_{{j-1}}}_{t_1}\mathrm{ant}_{0\mapsto 1}(\tilde{t}_{j})\,\d \tilde{t}_{j}\,,
\label{eq:dPm0}
\end{equation}
where we have used $\tilde{t}$ to denote the (sequentially nested) veto scales and we have suppressed any dependence on further phase-space variables to avoid clutter. With 
the initial condition $\tilde{t}_{{0}} = t_0$, the boundaries in the nested 
$\d \tilde{t}_{j}$ integrals follow from the fact that the first veto scale can be anywhere between $t_0$ and $t_1$, the second can be anywhere between that of the first veto scale, $\tilde{t}_{1}$, and $t_1$, and so on. 

The general case, for an arbitrary number of vetoes at each branching step, is:
\begin{equation}
\frac{\d {\cal P}_{m_1m_2\ldots m_n}}{\d \Phi_n} ~=~\frac{\d {\cal P}_{00\ldots 0}}{\d \Phi_n} 
\prod_{k=1}^n\prod_{j=1}^{m_k}\int^{\tilde{t}_{k_{j-1}}}_{t_k}\mathrm{ant}_{k-1\mapsto k}(\tilde{t}_{k_j})\,\d \tilde{t}_{k_j}
\,,
\label{eq:dPmm}
\end{equation}
with the boundary conditions $\tilde{t}_{k_0} = t_{k-1}$.

Nested integrals of the type appearing in eqs.~(\ref{eq:dPm0}) and (\ref{eq:dPmm}) can be recognised as the integral of a product over the ordered hypertriangle of the hypercube (see, e.g.,~\cite{Sjostrand:1987su}),
\begin{equation}
\prod_{j=1}^m\int^{\tilde{t}_{j-1}}_{t}\mathrm{ant}_i(\tilde{t}_j)\,\d \tilde{t}_j ~=~ \frac{1}{m!} \left( 
\int^{\tilde{t}_0}_{t}\mathrm{ant}_{i}(\tilde{t}_j)\,\d \tilde{t}_j \right)^m\,,\label{eq:nested}
\end{equation}
so that summing over all $m$ (now including $m=0$) produces an exponential,
\begin{equation}
\sum_{m=0}^\infty \prod_{j=1}^m\int^{\tilde{t}_{j-1}}_{t}\mathrm{ant}_{i}(\tilde{t}_j)\,\d \tilde{t}_j~=~ \exp\left(\int^{\tilde{t}_{0}}_{t}\mathrm{ant}_{i}(\tilde{t}_j)\,\d \tilde{t}_j\right)\,.
\label{eq:exp}
\end{equation}
This is simply the inverse of a corresponding Sudakov factor, $\Delta_i(\tilde{t}_0,t)$. Extending the sums over $j$ in \eqRef{eq:dPmm} to all $m_k\ge 0$, therefore yields a product of inverse Sudakov factors, 
\begin{equation}
\sum_{m_k\ge0}\frac{\d {\cal P}_{m_1m_2\ldots m_n}}{\d \Phi_n} ~=~\frac{\d {\cal P}_{00\ldots 0}}{\d \Phi_n} 
\prod_{k=1}^n\,
\frac{1}{\Delta_k(t_{k-1},t_k)}\,,
\label{eq:invSud}
\end{equation}
which precisely cancel those in \eqRef{eq:dP00}. The left-hand side of \eqRef{eq:invSud} is the phase-space sampling generated by the  \Sunshine algorithm. 
This proves that \Sunshine produces events distributed according to the product of shower kernels given in 
\eqRef{eq:tree}, i.e., that 
\begin{equation}
\hspace*{-3cm}\mbox{\Sunshine~:~~~}\sum_{m_k\ge 0}\frac{\d {\cal P}_{m_1m_2\ldots m_n}}{\d \Phi_n} ~=~|M_0|^2 \prod_{i=0}^{n-1}\mathrm{ant}_{i\mapsto i+1}\,.
\end{equation}
This is what we shall exploit to study tree-level shower expansions in sec.~\ref{sec:treeShower}.

With MECs, the generated distribution is proportional to the corresponding tree-level ME:
\begin{equation}
\frac{\d {\cal P}}{\d \Phi_{n}} = |M_n|^2\,.
\end{equation}
The proof is completely analogous to the one above so we do not repeat it. 

One subtlety is that the above proof does not show explicitly what happens if one uses so-called power showers~\cite{Plehn:2005cq}, smooth ordering~\cite{Corke:2010zj,Giele:2011cb,Hartgring:2013jma}, or a combination of $2\mapsto 3$ and $2\mapsto 4$ antennae~\cite{Li:2016yez,Campbell:2021svd,El-Menoufi:2024sys} to ensure full phase-space coverage. This is addressed in the next subsection.

\subsection{Ensuring full phase-space coverage}
\label{sec:fullPS}

Strictly speaking the proof in sec.~\ref{sec:newAlgo} only shows that a sampling proportional to the tree-level expansion is obtained \emph{for phase-space points which are reached by the shower}.  In general, however, the requirement of strong ordering of the shower evolution scales may produce ``dead zones'', points which cannot be reached by any ordered sequence of branchings. Whether this occurs or not for a given algorithm depends on its choices for evolution variable(s) and kinematics map(s). For algorithms that do exhibit this issue (such as the sector shower we have chosen to focus on in this work~\cite{Brooks:2020upa}), we here comment on three approaches to overcome it: power showers~\cite{Plehn:2005cq}, smooth ordering~\cite{Giele:2011cb}, and $2\mapsto N$ branchings~\cite{Li:2016yez}. 

Fig.~\ref{fig:2to4} illustrates the main difference between the two former approaches (power showers and smooth ordering) and the latter one ($2\mapsto N$ branchings), for a parton configuration (C) which we assume cannot be reached by any ordered sequence of $n\mapsto n+1$ branchings: 
\begin{figure}[t]
\centering
\includegraphics*[width=0.58\textwidth]{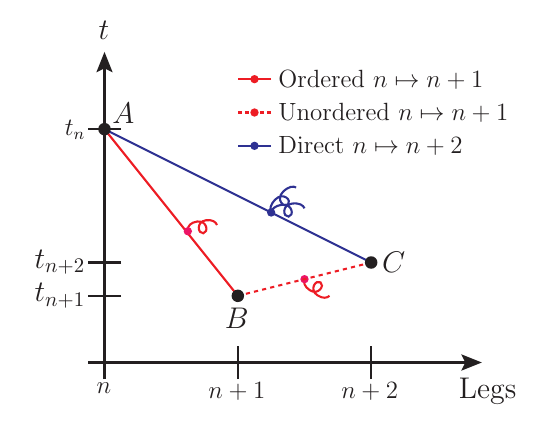}
\caption{\label{fig:2to4}The case of an ($n+2$)-parton configuration, $C$, which can only be reached through unordered $n\mapsto n+1$ branchings (red lines) --- e.g., via the on-shell ($n+1$)-parton configuration at $B$. Alternatively,  in a shower that allows for direct $n \mapsto n+2$ branchings the $A\mapsto C$ transition can be generated directly (blue line), bypassing the configuration at $B$.}
\end{figure}
\begin{enumerate}[label=\Alph*)]
\item An $n$-parton configuration defined at the scale $t_n$ starts evolving. 
\item An intermediate on-shell $(n+1)$-parton configuration with $t_{n+1}<t_n$ can be reached by a strongly-ordered $n\mapsto n+1$ branching (solid red line). This configuration would be reachable by any of the showers discussed here.
\item This particular parton configuration is assumed to be inaccessible to a conventional strongly-ordered shower that only includes $n\mapsto n+1$ branchings. \begin{itemize}
    \item 
The red dotted line represents an unordered $n\mapsto n+1$ branching, with power showers and smooth ordering merely differing by whether the combined branching kernel is assumed to be proportional to $\alpha_s(t_{n+1})\alpha_s(t_{n+2})/(t_{n+1} t_{n+2})$ or to $\alpha_s^2(t_{n+2})/t_{n+2}^2$, respectively (in the limit $t_{n+2} \gg t_{n+1}$). \item The blue solid line illustrates a direct $n\to n+2$ branching. This has a similar combined branching kernel as smooth ordering ($\propto \alpha_s^2(t_{n+2})/t_{n+2}^2)$ but since it bypasses the phase-space point at B, the associated Sudakov factor does not depend on $t_{n+1}$ at all, while both power showers and smooth ordering will include a factor $\Delta(t_n,t_{n+1})$ in their combined Sudakov factors for $A\mapsto C$. 
\end{itemize}
\end{enumerate}

\subsubsection*{Power showers and smooth ordering}
The simplest prescription for covering all of phase space for hard branchings is to use the phase-space maximum of each antenna as the restart scale after each accepted branching, instead of the scale of the branching. In its simplest incarnation (power showers~\cite{Plehn:2005cq}), this tends to significantly overestimate the actual matrix elements~\cite{Plehn:2005cq,Giele:2011cb}. Nonetheless, if the aim is merely to ensure a complete population of the post-branching phase space, then this approach is guaranteed to do that. In the context of \Sunshine, the price one pays is a comparative overrepresentation of configurations that can only be reached through unordered sequences of branching scales, $t_{n>m} > t_m$, such as the one illustrated in \figRef{fig:2to4}. A matrix-element correction (or reweighting) would therefore be expected to have relatively low efficiency for such configurations but since they typically only occupy a small fraction of phase space (of order a few percent~\cite{Giele:2011cb}) and do not exhibit any leading-logarithmic enhancements, this may not be a significant showstopper in terms of overall sampling efficiency. The proof that the \Sunshine algorithm still eliminates Sudakov factors for power showers is as follows. 

The sample produced by following all of the branchings that would normally be generated by a power shower produces the same distribution as in \eqRef{eq:ps} but with the starting scale of each of the Sudakov factors replaced by the evolution-variable maximum for each $n\mapsto n +1$ stage. Since this is always bounded from above by $t_0$, the power-shower equivalent of \eqRef{eq:dP00} can be expressed as:
\begin{equation}
\frac{\d {\cal P}^{\mathrm{power}}_{00\ldots 0}}{\d \Phi_n} = |M_0|^2 \prod_{i=0}^{n-1} \mathrm{ant}_{i\to i+1}\Delta_i(t_0, t_{i+1})~\,.\label{eq:power}
\end{equation}
For the distribution of \Sunshine events with an arbitrary number of vetoes at each step, \eqRef{eq:dPmm}, only the boundary condition changes, from $\tilde{t}_{k_0} = t_{k-1}$ to $\tilde{t}_{k_0} = t_0$. In \eqsRef{eq:nested} and \eqref{eq:exp}, the only change is that $\tilde{t}_0$ is replaced by $t_0$, which thus also becomes the first argument of all of the Sudakov factors in the denominator of \eqRef{eq:invSud}. Since these  exactly cancel those in \eqRef{eq:power}, we have proved that also in the power-shower mode, the \Sunshine algorithm eliminates the Sudakov factors. 

A better approximation than power showers is obtained by applying a dampening factor interpolating between a $1/p_\perp^2$ below and a $1/p_\perp^4$ behaviour above the scale of the preceding branching~\cite{Corke:2010zj,Giele:2011cb}. This is equivalent to replacing the strong-ordering step function of a conventional shower by a smooth dampening factor~\cite{Giele:2011cb},
\begin{equation}
\Theta\big(t_{n}-t_{n+1}\big) \,\to\, \frac{t_n}{t_n + t_{n+1}}\,.
\end{equation}
Thus, the smooth-ordering antenna functions and Sudakov factors are, respectively: 
\begin{eqnarray}
\mathrm{ant}_{i\mapsto i+1}^{\mathrm{smooth}} & ~=~ & 
  \frac{t_i}{t_i+t_{i+1}}\mathrm{ant}_{i\mapsto i+1} \\
  \Delta_i^{\mathrm{smooth}}(t_0, t_i, t_{i+1}) & = &
  \exp\Big( 
  -  \int_{t_{i+1}}^{t_0} \frac{t_i}{t_i+\tilde{t}}\, 
  \mathrm{ant}_{i\mapsto i+1} (\tilde{t}) \,\d\tilde{t}
\Big)\,.
\end{eqnarray}
The phase-space limits are the same as for power showers and hence the proof that the \Sunshine algorithm cancels the Sudakov factors is the same as for power showers.

\subsubsection*{$\mathbf{2\mapsto N}$ branchings}

An alternative approach to covering the full $N$-parton phase space(s) is that of so-called ``direct'' $2\mapsto N$ branchings~\cite{Li:2016yez}. This method exploits the same exact nested on-shell phase-space factorisations~\cite{Kosower:2003bh,Giele:2007di} as conventional $2\mapsto 3$ antenna showers do, but any $N$-parton configuration that can only be reached via an unordered sequence of $2\mapsto 3$ branchings is generated ``directly'', as a single $2\mapsto N$ branching. This is achieved by effectively integrating out the scale(s) of the intermediate unordered phase-space point(s).  

In the original study~\cite{Li:2016yez}, this method was only worked out for $2\mapsto 4$ branchings, hence also here we shall restrict our attention to this case. To reach parton multiplicities beyond 4 partons, the current implementation of the \Sunshine algorithm therefore reverts to smooth ordering (or power showers) for configurations that cannot be reached by any direct ($2\mapsto 3$ or $2\mapsto 4$) branching. 

For phase-space configurations that \emph{can} be reached via sequences of $2\mapsto3$ branchings, the product of antenna functions, $\mathrm{ant}_{i\mapsto i+1}$, remains unchanged. However, the Sudakov factor for the Born-level evolution must take the addition of the direct $2\mapsto 4$ branchings into account,
\begin{equation}
\Delta_0(t_0, t) ~=~ \Delta_{0\mapsto 1}(t_0,t)\,\Delta_{0\mapsto 2}(t_0, t)\,.
\end{equation}
In the \Sunshine algorithm, each $2\mapsto 4$ branching will now also lead to a bifurcation of the event evolution, with one event in which the branching is accepted and one in which it is rejected. This leads to factors of 
\begin{equation}
\int^{t_0}_{t} \mathrm{ant}_{0\mapsto 2}(\tilde{t})\, \d \tilde{t}\,,
\end{equation}
which are completely analogous to those for $\mathrm{ant}_{0\mapsto 1}$ already covered. The proof that these factors sum to produce the exact inverse of $\Delta_{0\mapsto 2}$ is identical to those already given and hence is not repeated here. 

For phase-space configurations that \emph{cannot} be reached via sequences of $2\mapsto 3$ branchings, the product of antenna functions starts off with a single factor of $\mathrm{ant}_{0\mapsto 2}$ and then reverts to those discussed above. From the point of view of the \Sunshine algorithm this case again presents no particular additional subtleties relative to those already covered. 

Probability-wise we therefore also see no conceptual issues with extending the \Sunshine algorithm to higher-multiplicity $2\mapsto N$ branchings if and when such become available.

\subsection{Validation tests}
\label{sec:sudakov-tests}

In the following, we aim to validate that the \Sunshine algorithm per the implementation in \textsc{Pythia} does indeed behave as expected — i.e. that the produced $n$-parton events generated by \Sunshine are equivalent to the product of shower kernels as per eq.~\eqref{eq:tree}.
In order to test this, we reintroduce the cancelled-out Sudakov factors back into \Sunshine via reweightings of \Sunshine events, and compare consistency of the Sudakov-weighted \Sunshine events to the physical event produced by conventional showers (i.e. the event given all generated branchings are accepted).
These missing Sudakov factors which we wish to reintroduce represent the no-emission probabilities between the branching scales of the \Sunshine event, analogous to those that are constructed in CKKW-L type merging strategies~\cite{Catani:2001cc,Lonnblad:2001iq} adapted to sector showers~\cite{Brooks:2020mab}. 
Thus for each $n\mapsto n+1$ branching that occurs in a \Sunshine event, we aim to re-impose the survival probability of the $n$-parton state up to the scale at which the $n\mapsto n+1$ branching occurs (generalisations including $2\mapsto 4$ mappings are elaborated on below).
To do so, a ``trial''-shower method~\cite{Lonnblad:2001iq} is used, with the prescription outlined below.
For distinction of notation, here $t_i$ will be used to denote the emission scale of the \Sunshine event in question, and $t'_i$ will be the scale from the trial shower.

First, we take the Born-level configuration of the \Sunshine event and perform a trial emission which occurs at some scale $t'_1$. If the trial emission occurs at a lower scale than that of the \Sunshine branching — i.e. $t'_1 < t_1$ — then the survival probability of the 2-parton state up to $t_1$ as estimated by the trial shower is zero, and thus we apply a zero weight to the \Sunshine event.
Should the trial emission scale be greater than for the \Sunshine event, $t'_1 > t_1$, then the event is kept with weight unity. 
This thereby reintroduces the missing Sudakov $\Delta_{0\mapsto 1} (t_0,t_1)$ into the \Sunshine event. 
This can be repeated on the 3-parton configuration of the \Sunshine event, with the trial shower then starting from the corresponding scale $t_1$,\footnote{Or from $t_0$ if using power showers.} now with the veto condition $t'_2 < t_2$, thus reconstructing the Sudakov $\Delta_{1\mapsto 2} (t_1,t_2)$. 
The process is repeated for all branchings of a given \Sunshine event, resulting in a weighting of unity if all trial emission scales result in $t'_i > t_i$, otherwise the event is vetoed with weight zero. 

For the inclusion of direct $2\mapsto 4$ mappings, the Sudakov we need to reconstruct is now $\Delta_{0\mapsto 1} (t_0,t_1)$ $\Delta_{0\mapsto 2} (t_0,t_1)$. 
The same prescription as above is followed for the Sudakov reweighting, with the following caveats: should the trial shower generate a direct $2\mapsto 4$ branching in contrast to a \Sunshine event with successive $2\mapsto 3\mapsto 4$ final states, the survival criterion for the \Sunshine event is then $t_2' > t_1$. Similarly, should the trial shower generate a $2\mapsto3$ branching whilst the \Sunshine event involves a direct $2\mapsto4$ branching at scale $t_2$, then the survival criterion becomes $t_1' > t_2$. Indeed, the Sudakov we aim to reproduce represents the survival probability of the 2-parton state up to the relevant scale, regardless of whether the 2-parton state is destroyed by a $2\mapsto3$ or a $2\mapsto4$ mapping.

\begin{figure}[tp]
    \centering
    \includegraphics[width=0.49\textwidth,page=1]{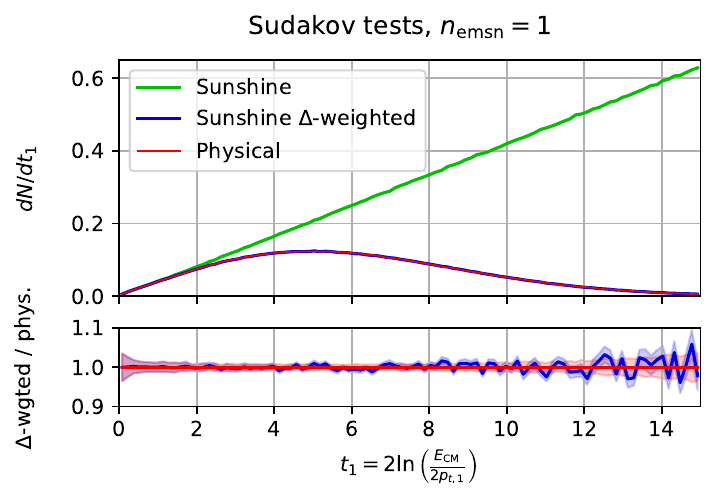}
    \includegraphics[width=0.49\textwidth,page=2]{figures/plot-test10.pdf}\\
    \includegraphics[width=0.49\textwidth,page=3]{figures/plot-test10.pdf}
    \includegraphics[width=0.49\textwidth,page=4]{figures/plot-test10.pdf}
    \caption{
    Validation tests of the \Sunshine algorithm, for one, two, three and four emissions. The plots show the distribution of the scale of the last branching, $t=2\ln E_{\mathrm{CM}}/2p_t$, according to \Sunshine (i.e.~a fixed-order expansion of the \Vincia shower, green), the Sudakov-weighted distribution of \Sunshine events (following the procedure in the text, blue) and the physical distribution from the shower truncated at the corresponding order (red). The Sudakov-weighted distribution agrees with the physical result in all cases, as shown in the ratio plot.
    }
    \label{fig:sudakov-tests}
\end{figure}

With the re-introduction of the Sudakov factors, the reweighted \Sunshine events should be consistent with the physical shower (i.e.~events where all the branchings have been accepted). This is shown in fig.~\ref{fig:sudakov-tests}, at first, second, third and fourth order. In all cases, we include matrix-element corrections up to the first two emissions, and the $2\mapsto 4$ branchings from the Born to the Born$+2$ state. (This effectively tests the interplay between the sectorised $2\mapsto 3$ and $2\mapsto 4$ branchings.)
Fig.~\ref{fig:sudakov-tests} displays the distribution of the scale of the last branching, normalised to the total energy, $t_i = 2\ln \frac{E_\mathrm{CM}}{2 p_{t,i}}$, for the pure \Sunshine sample (green), the reweighted \Sunshine events (blue), and for the physical shower (red). The distributions of the Sudakov-weighted \Sunshine events and the physical showers agree, as expected.

\subsection{Speed of the \textsc{Sunshine} algorithm}
\label{sec:speed}

The main factors affecting the rate of events generated by the \Sunshine algorithm per unit time are:
\begin{itemize}
\item The speed of the underlying shower algorithm. 
\item The number of branchings requested from the shower.
\item The value of the IR cutoff. 
\item The speed and efficiency of any (exact or approximate) matrix-element corrections applied during the shower generation. 
\item The speed and efficiency of any post-hoc unweighting step applied to the final events.
\end{itemize}

\begin{figure}[t]
\centering
\includegraphics*[width=0.48\textwidth]{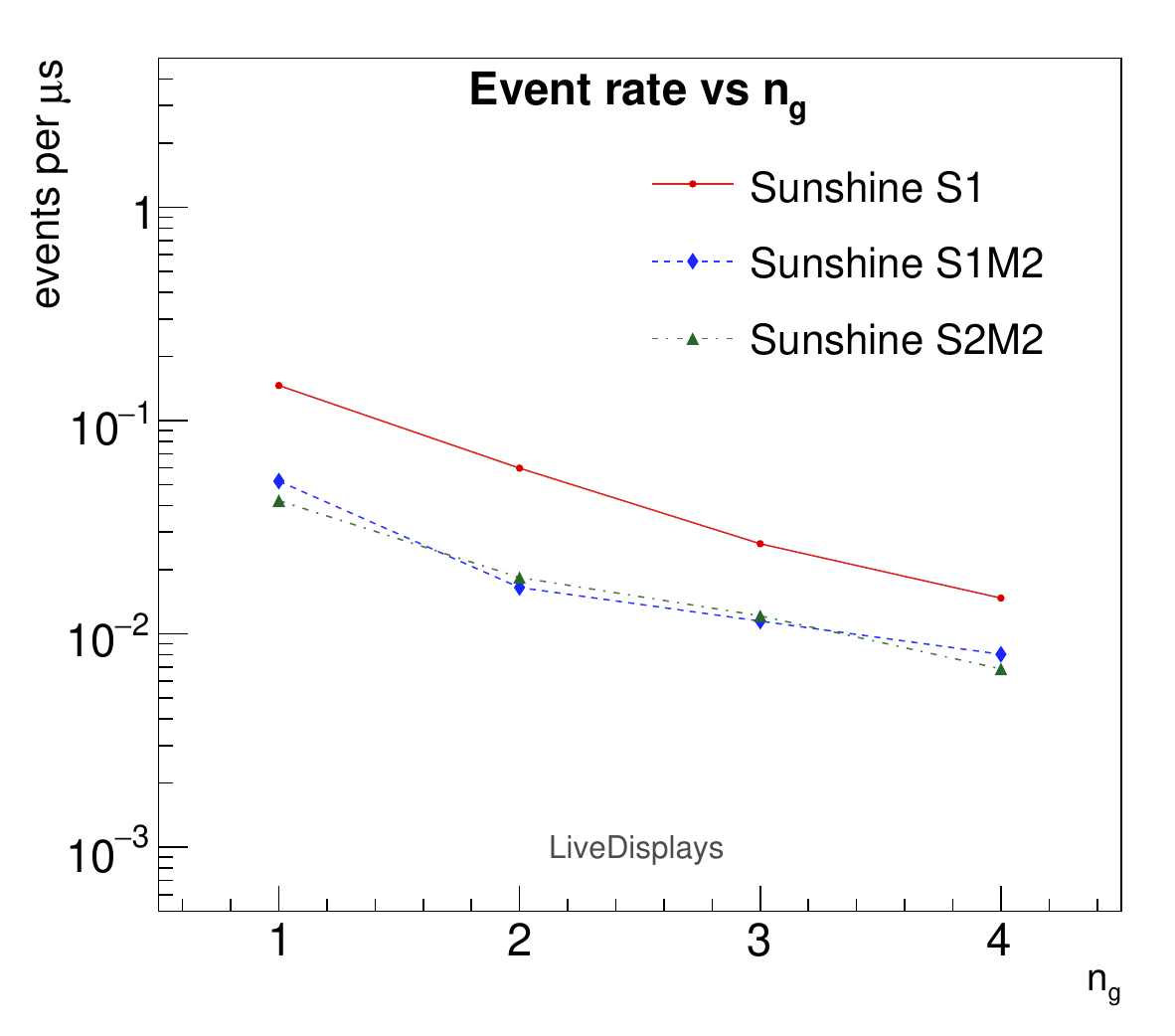}
\caption{Event rate (per microsecond) as a function of the number of emitted gluons $n_g$, for the \Sunshine algorithm with a pure $2\mapsto 3$ shower (S1, solid red line), a $2\mapsto 3$ shower incorporating matrix-element corrections through $n_g \le 2$ (S1M2, dashed blue line), and for a shower that combines $2\mapsto 3$ and $2\mapsto 4$ branchings matched through $n_g \le   2$ (S2M2, dot-dashed green line).}
\label{fig:rateVng}
\end{figure}
An illustration of the two first points is given by the {solid red line} in \figRef{fig:rateVng}, for our reference case of the \Vincia sector shower as the baseline shower, with fixed $\alpha_s = 0.12$, starting scale $t_0 = M_Z^2$, IR cutoff scale $t_\mathrm{cut} = (0.5\,\mathrm{GeV})^2$, and using power showers to fill phase space. The vertical axis shows events per microsecond (on a single-threaded Apple M2 Pro) and the horizontal axis shows the number of requested branchings. (For simplicity, only gluon emissions were allowed in these runs so $n_g$ counts the number of branchings generated by the algorithm.) 

The legend label S1 for the solid red line means that the shower algorithm was run with only $n\mapsto n+1$ branchings (i.e., no $2\mapsto 4$ branchings) and without matrix-element corrections. The resulting phase-space distribution is thus simply proportional to the Born-level matrix element times a product of $2\mapsto 3$ antenna functions. For these conditions, the speed of the algorithm starts at just over 0.1 event per microsecond for one branching, dropping by about a factor 2.2 per extra branching. 

The dashed blue line in \figRef{fig:rateVng} illustrates the same conditions but imposing matrix-element corrections for up to two gluons (indicated by M2 in the legend label). This implies that the phase-space population obtained for up to 2 branchings is simply proportional to the corresponding tree-level matrix element, while for 3 and higher branchings it is proportional to the 4-parton matrix element times a (product of) $2\mapsto 3$ antenna functions. The drop in speed by a factor $\sim 5$ is mainly caused by the fact that the imposition of MECs during the shower forces one to use larger overestimates in the shower veto algorithm, here done rather crudely by simply applying a multiplicative factor of 3.5 to all ME-corrected branchings. Note that, since this extra overestimating factor is not applied beyond 2 gluons (in this example), the speed penalty reduces from that point onwards. 

Finally, the dot-dashed green line illustrates the speed if one uses $2\mapsto 4$ branchings to fill the 4-parton phase space instead of $2\mapsto 3$ power showers. Since some events now go directly from 0 to 2 gluons, the speed for $n_g = 1$ reduces slightly, while that for $n_g = 2$ increases slightly, but these differences are relatively minor. (The slight difference at $n_g = 4$ arises because the $2\mapsto 4$ shower currently defaults back to ordered showers, rather than power showers, beyond $n_g = 2$.)

\begin{figure}[t]
\centering
\includegraphics*[width=0.48\textwidth]{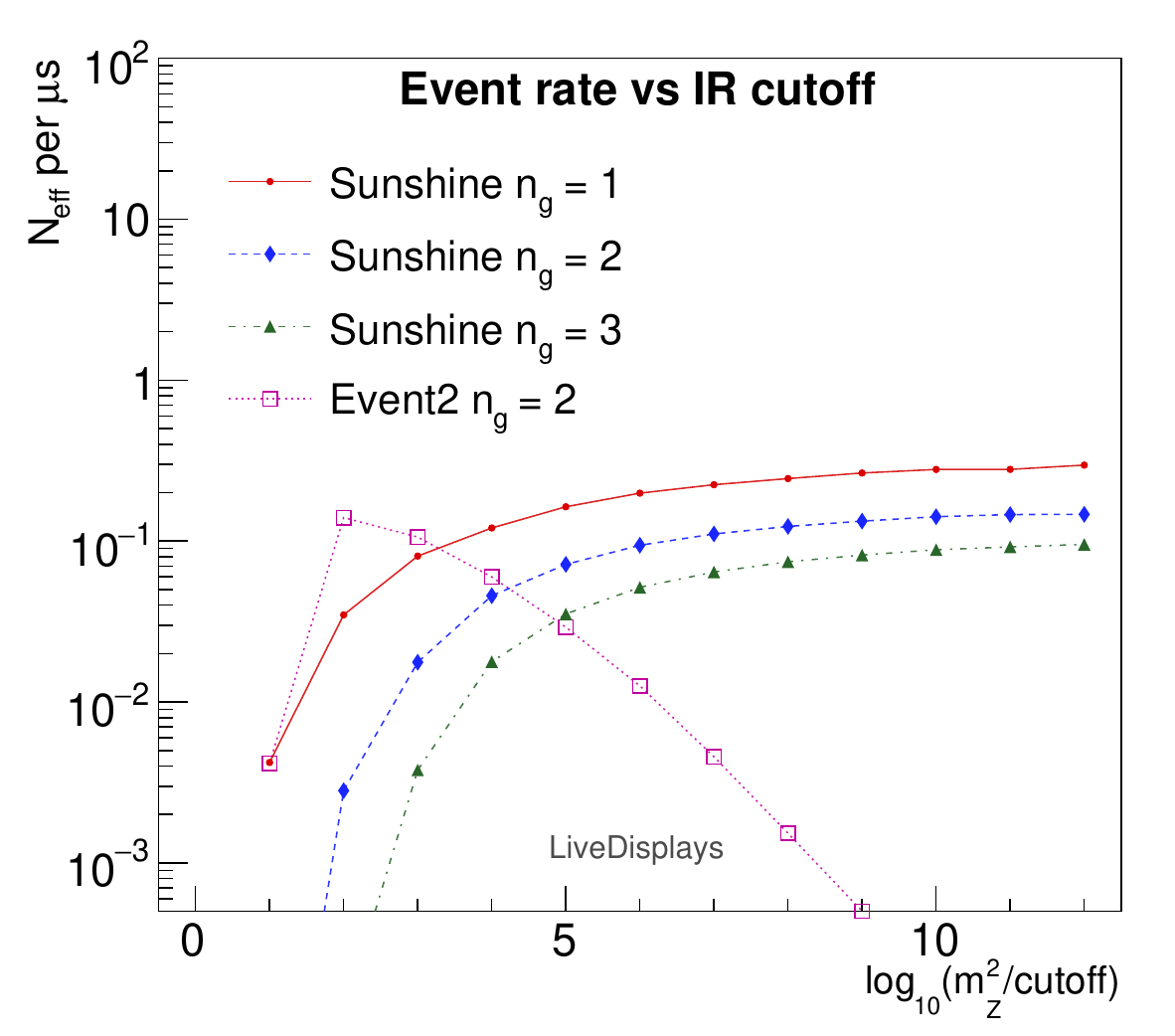}
\caption{Effective event rate (per microsecond) as function of the infrared cutoff, for \Sunshine with $n_g = 1$, $2$, and $3$, and for \EventTwo with $n_g = 2$. The settings for \Sunshine are the same as for S1 in \figRef{fig:rateVng}; those for \EventTwo are described in the text.}
\label{fig:rateVcut}
\end{figure}
\FigRef{fig:rateVcut} shows the event rate as one varies the IR cutoff, for {$n_g = 1$} {(solid red line)}, $n_g=2$ {(dashed blue line)}, and $n_g=3$ {(dot-dashed green line)}, with the S1 setup (power showers without direct $2\mapsto4$ branchings and no MECs). Also shown is 
the event-generation rate obtained for 4-parton events (i.e., $n_g = 2$) with  the dedicated fixed-order generator {\EventTwo}~\cite{Catani:1996jh,Catani:1996vz} {(dotted purple line)}. For as fair a comparison as possible, the latter was run in a special mode with the 2- and 3-parton virtual corrections disabled, and since it outputs weighted events the rate plotted for \EventTwo is $N_\mathrm{eff} = (\sum w)^2 / \sum w^2$. (The corresponding rate of fully-unweighted events is one to two orders of magnitude smaller.) We note also that the IR cutoffs in \Vincia and in \EventTwo are not defined in exactly the same way. In the former, the cutoff is taken in the $p_\perp$ evolution scale, while in the latter it is taken in invariant mass. These definitions coincide in the hard-collinear limit but not in the soft limit where the \Vincia variable vanishes quadratically while the \EventTwo one only goes linearly to zero. Because of these subtleties (shower-unweighted vs ME-weighted events and cutoff mismatches), one can only draw qualitative conclusions from the comparison in \figRef{fig:rateVcut}. The obvious such is that the \Sunshine algorithm asymptotes to constant rates for small cutoff values (towards the right edge of the plot), while the corresponding rate for \EventTwo decreases. 

At the left edge of the plot, one sees that \EventTwo does outdo \Vincia\!+\Sunshine for large values of the cutoff, above $\sim 10^{-4}M_Z^2 \sim (1\,\mathrm{GeV})^2$. This also makes sense: by construction, \Sunshine is most efficient in the singular regions of phase space.

\begin{figure}[t]
\centering
\includegraphics*[width=0.48\textwidth]{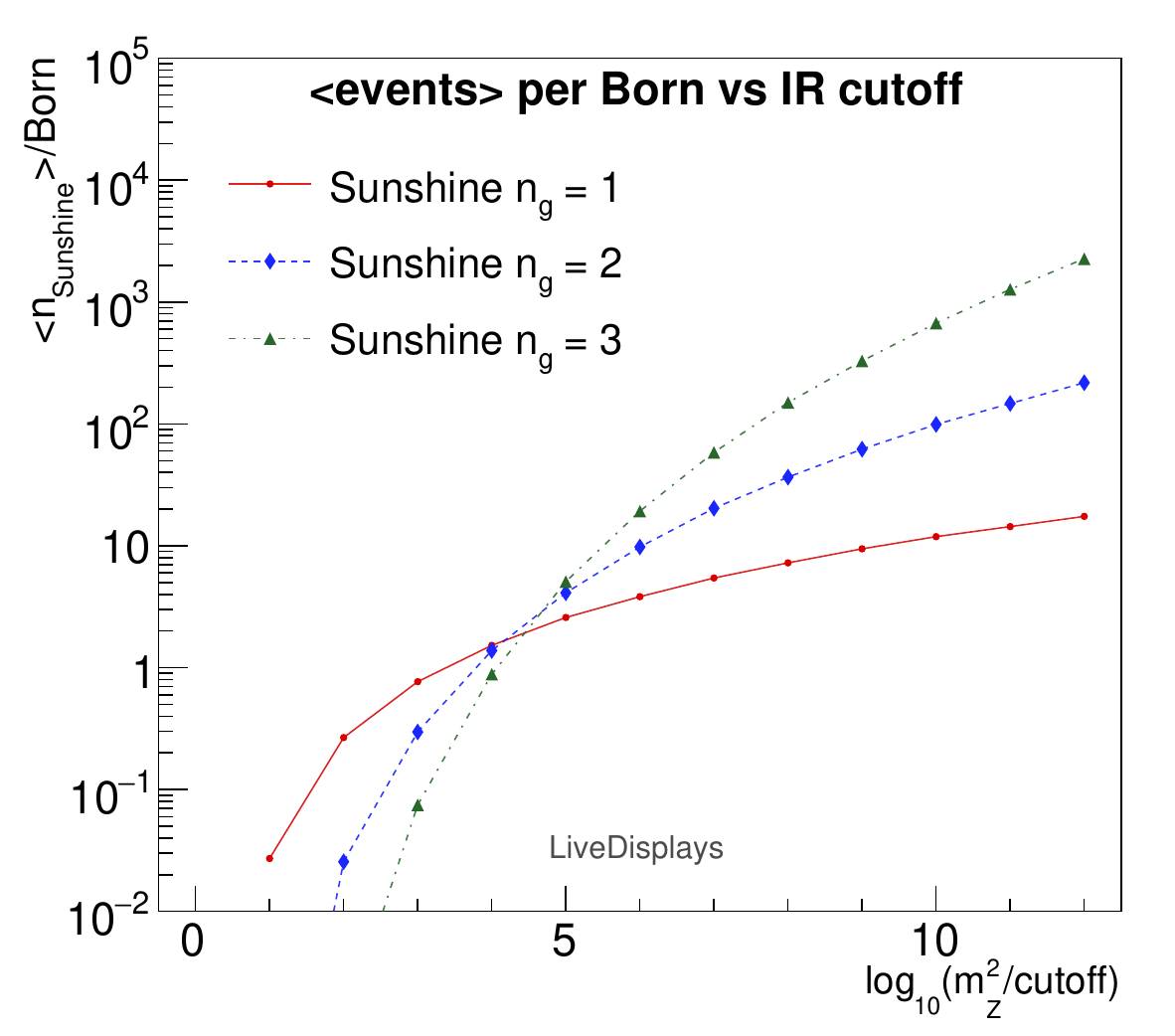}\
\caption{The average number of branched events with $n_g = 1$, $2$, and $3$, produced per Born event by the \Sunshine algorithm (with settings as for S1 in \figRef{fig:rateVng}), as a function of the shower IR cutoff.}
\label{fig:multVcut}
\end{figure}
The comparatively high event-generation rate of \Sunshine for small cutoff values can be further elucidated by the plot shown in \figRef{fig:multVcut}, which shows the average number of events with $1$, $2$, and $3$ branchings produced by the \Sunshine algorithm \emph{per} Born event, as a function of the IR cutoff. For an ordinary (unitary) shower, this number would always be smaller than unity.\footnote{Specifically, it would be equal to one minus the sum of exclusive fractions with less than $n_g$ branchings.} For large cutoffs (towards the left edge of the plot), one does indeed see that \Sunshine produces an average of less than one branched event per Born event, reflecting that the shower only produces very hard branchings in a small fraction of events. But towards the right-hand edge of the figure, the unitarity violation comes into its own, with each Born event acting as the seed for tens, hundreds, or even thousands of branched events. This translates to very high sampling statistics in the IR regions of phase space. 

\subsection{Comments on numerical stability}
\label{sec:stability}

Most modern shower algorithms are formulated in terms of double-precision floating-point arithmetic. In the context of parton-shower Monte Carlos, for which the overall theoretical uncertainty is never below ${\cal O}(1\%)$, this is typically sufficient for present-day collider physics studies. Exceptions are problems involving extreme scale ratios such as ultra-high-energy cosmic-ray interactions, heavy dark-matter annihilation~\cite{Bauer:2020jay,Brooks:2021kji}, and numerical tests of the logarithmic accuracy of parton showers for asymptotically large scale  hierarchies~\cite{Dasgupta:2020fwr,Hamilton:2020rcu,Karlberg:2021kwr,Hamilton:2021dyz,vanBeekveld:2022zhl,vanBeekveld:2022ukn,Herren:2022jej,vanBeekveld:2023chs,vanBeekveld:2024wws,vanBeekveld:2025lpz}.
Such studies have already led to the development of methods that can reach higher numerical accuracy, e.g.~in the \textsc{PanScales} showers, which combine the use of higher-precision numerical libraries such as MPFR~\cite{MPFR} and QD~\cite{hida2000quad}, an in-house \texttt{doubleexp} type~\cite{vanBeekveld:2023ivn}, and dedicated representations of kinematics and Lorentz transformations. 

At the time of writing however, such improvements have not yet been implemented in the \Vincia sector shower, whose effective dynamical range therefore remains limited by generic double-precision numerical stability. This range 
is illustrated by the plots in \figsRef{fig:rateVcut} and \ref{fig:multVcut} in the preceding subsection, whose horizontal axes span 12 orders of magnitude in squared invariants, or 6 in linear ones. (A similar range of numerical stability was found for \Vincia's EW showers in heavy dark-matter annihilation~\cite{Brooks:2021kji}.)
Exploring larger hierarchies than this would require improvements to \Vincia analogous to those developed for the \textsc{PanScales} code. We do not regard these as essential to present the \Sunshine algorithm to a broader audience but do intend to return to this in a future study. 

\section{Tree-level tests of parton showers \label{sec:treeShower}}
Finally, we present an example of a useful application of the algorithm of sec.~\ref{sec:newAlgo}, namely in the context of tree-level matrix-element tests of parton showers. Here we specifically apply \Sunshine to the \Vincia sector shower~\cite{Brooks:2020upa} with nested matrix-element corrections (MEC).

In parton showers, MECs are imposed multiplicatively in the form of an extra acceptance probability factor, within the Sudakov veto algorithm: they can be included in a nested manner, order-by-order, and are designed to improve the shower description of hard radiation. Additionally, the real-emission correction is also a necessary ingredient in fixed-order matching schemes that use the shower as a phase-space generator for the matched emissions. With MECs enabled, the tree-level expansion of the shower at order $\mathcal{O}(\alpha_s^n)$ is expected to reproduce the correct tree-level matrix element squared at the same order.

To our knowledge, the correctness of such MEC implementations in parton shower codes is typically tested with a varying degree of robustness, either by 1) computing separately, analytically or numerically, the form of eq.~\eqref{eq:ps} directly from the shower kinematic map and kernels, or 2) writing extra code that undergoes the same steps as in the shower decision tree, computes the various acceptance factors, and returns a product of these instead of actually accepting or rejecting emissions (as in the public version of the \textsc{PanScales} showers~\cite{vanBeekveld:2023ivn}).\footnote{In \textsc{PanScales}, the matching accuracy was also recently tested by running the shower at several finite values of $\alpha_s$, and extracting the NLO coefficient by extrapolating $\alpha_s \to 0$~\cite{vanBeekveld:2023ivn}.}
Instead, using the algorithm of sec.~\ref{sec:newAlgo}, one can obtain \textit{automatically} the tree-level expansion of the shower at any given order, running the shower algorithm as is.
In the following we consider the simple case of hadronic decays in $e^+e^-$ collisions, $e^+e^- \to Z/\gamma^* \to q \bar q$, at first and second order, i.e.~up to $Z/\gamma^*\to 4$ jets, and examine the effect of nested MECs.

For all fixed-order tests below, we run the \Vincia sector shower at a centre-of-mass energy $\sqrt{s}=M_Z=91.2$ GeV, with a fixed value of the strong coupling, $\alpha_s(M_Z)=0.1$, and set all quark masses to zero. The shower infrared cutoff is set to $t_{\mathrm{cut}}=0.01$ GeV, and we run the shower purely perturbatively (i.e.~only the QCD shower, without hadronisation/MPI). Additionally, \Vincia turns off MECs by default when the scale of the emission is below a fixed cutoff, $k_t < t_{\mathrm{cut}}^\mathrm{match} = 2$ GeV: here we instead run with MECs enabled all the way down to the shower cutoff. Finally, the trial emission density needs to be enhanced for the MEC, 
\begin{equation}
    w_{n+1}^{\mathrm{MEC}} \,=\, \frac{|M_{n+1}|^2}{ \mathcal{H}^{\mathrm{MEC}} \, \mathrm{ant_{n \mapsto n+1}}\,|M_n^2|}\,,
\end{equation}
to be bounded by one, where $\mathcal{H}^{\mathrm{MEC}}=4$ is simply a multiplicative so-called ``headroom factor'' which is applied to the trial probability density to ensure $w^\mathrm{MEC} < 1$. 

Note that all prefactors, e.g.~Jacobian factors from the parton-shower variables to the physical phase space, colour factors, as well as powers of $\alpha_s/2\pi$, are automatically included in the \Sunshine output (since they also enter the acceptance probability at every stage of the shower veto algorithm), i.e.~the events produced by \Sunshine can be directly binned --- with unit weight --- to populate the fixed-order expansion of any observable.
As a final comment, when we show results with MECs disabled, we will use the \texttt{modeSLC=2} approximation of subleading-colour effects~\cite{VinciaQCDSettings} (since this is the default in \Vincia, and it is in closer agreement to the full result).\footnote{The scheme uses a factor $2C_F$ for $q\bar q$-antennae, $C_A$ for $gg$-antennae, and interpolates between the corresponding colour factors in each collinear limit for $qg$-antennae, \begin{equation}\mathcal{C}(y_{ij},y_{jk})=2C_F \frac{1-y_{ij}}{2-y_{ij}-y_{jk}} + C_A \frac{1-y_{jk}}{2-y_{ij}-y_{jk}}\,. \nonumber\end{equation}}

\subsection{First-order tests}
\label{sec:fo-tests-as1}

We start at relative order $\mathcal{O}(\alpha_s)$ from the Born process, i.e.~we consider the final state $Z/\gamma^* \to q\bar q g$, and
examine the distribution of the quark momentum fractions,
\begin{equation}
    x_{q/\bar q} = \frac{2 p_{q/\bar q} \cdot Q}{Q^2}\,,
\end{equation}
in the two-dimensional Dalitz plane. The exact matrix element (integrated over Euler angles) is given by
\begin{equation}
    d\mathcal{P}_\text{exact} = \frac{1}{\sigma_B}d\sigma = dx_q dx_{\bar q}\frac{\alpha_s C_F}{2\pi}\frac{x_q^2 + x_{\bar q}^2}{(1-x_q)(1-x_{\bar q})}\,.
\label{eq:exact-me-as1}
\end{equation}
Note that \Vincia uses a final-final quark-antiquark antenna (stripped of colour and coupling factors) that differs from the exact matrix element squared ($A_3^0$, in the notation of Ref.~\cite{Gehrmann-DeRidder:2005btv}),
\begin{equation}
 \mathrm{ant}_{q_I \bar q_K \to q_i g_j \bar q_k} ~\equiv~ A_3^0(1_q, 3_g, 2_{\bar q}) \,+\, \frac{1}{s_{IK}}\,,
\end{equation}
by a non-singular term (it is effectively the average of $Z$ and $H$ decays)~\cite{Brooks:2020upa}.
\begin{figure}[tp]
    \centering
    \includegraphics[width=0.8\textwidth,page=1]{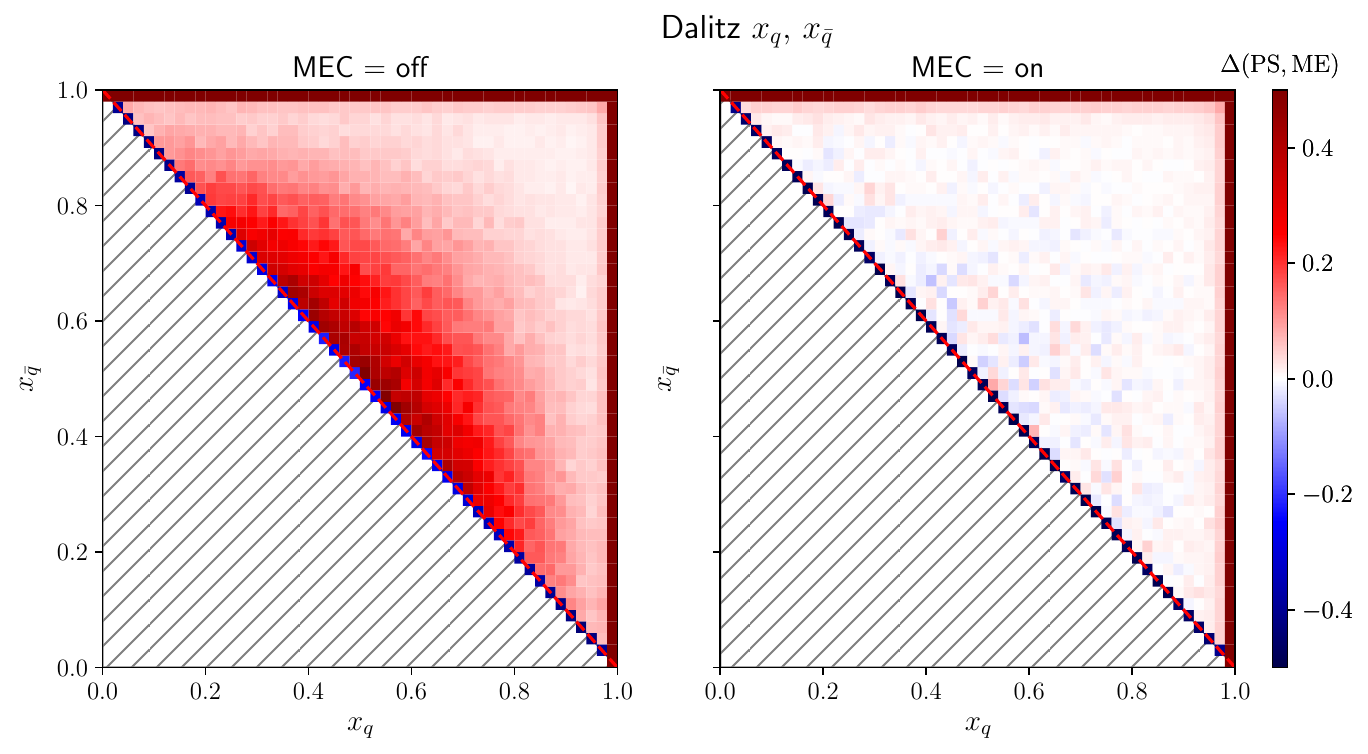}\\
    \includegraphics[width=0.8\textwidth,page=2]{figures/plot-test06-as1.pdf}
    \caption{The discrepancy $\Delta(\textrm{PS},\textrm{ME})$, as in eq.~\eqref{eq:delta-ps-me-as1}, between the \Vincia parton shower expanded at $\mathcal{O}(\alpha_s)$ with \Sunshine, to the exact matrix element from \eqRef{eq:exact-me-as1}. The ratio is plotted as a function of the Dalitz variables $x_q$ and $x_{\bar q}$ (top) and logarithmically (bottom). The relative discrepancy is shown on a colour scale, where white indicates agreement, without (left) and with (right) MECs.
     }
    \label{fig:fo-tests-as1-dalitz}
\end{figure}
Fig.~\ref{fig:fo-tests-as1-dalitz} shows the relative difference of the parton-shower effective tree-level matrix element, as provided by running \Vincia within the \Sunshine algorithm, to the exact matrix element of~\eqRef{eq:exact-me-as1},
\begin{equation}
    \Delta(\mathrm{PS},\mathrm{ME}) = \frac{d\mathcal{P}_\text{PS} - d\mathcal{P}_\text{exact}}{d\mathcal{P}_\text{exact}} = \frac{d\mathcal{P}_\text{PS}}{d\mathcal{P}_\text{exact}}-1\,,
    \label{eq:delta-ps-me-as1}
\end{equation}
i.e.~a value of zero indicates agreement with the exact tree-level matrix element. This is depicted with MECs disabled (left) and enabled (right), in linear $x_{q / \bar q}$ space (top) and logarithmic space $\ln ( 1- x_{q / \bar q})$ (bottom). One observes the uncorrected shower matrix element to depart from the exact result in the hard, large-angle region (i.e. $x_q+x_{\bar q} \to 1$), where the discrepancy is lifted once MECs are activated.

\begin{figure}[tp]
    \centering
    \!\includegraphics*[width=0.499\textwidth,page=3]{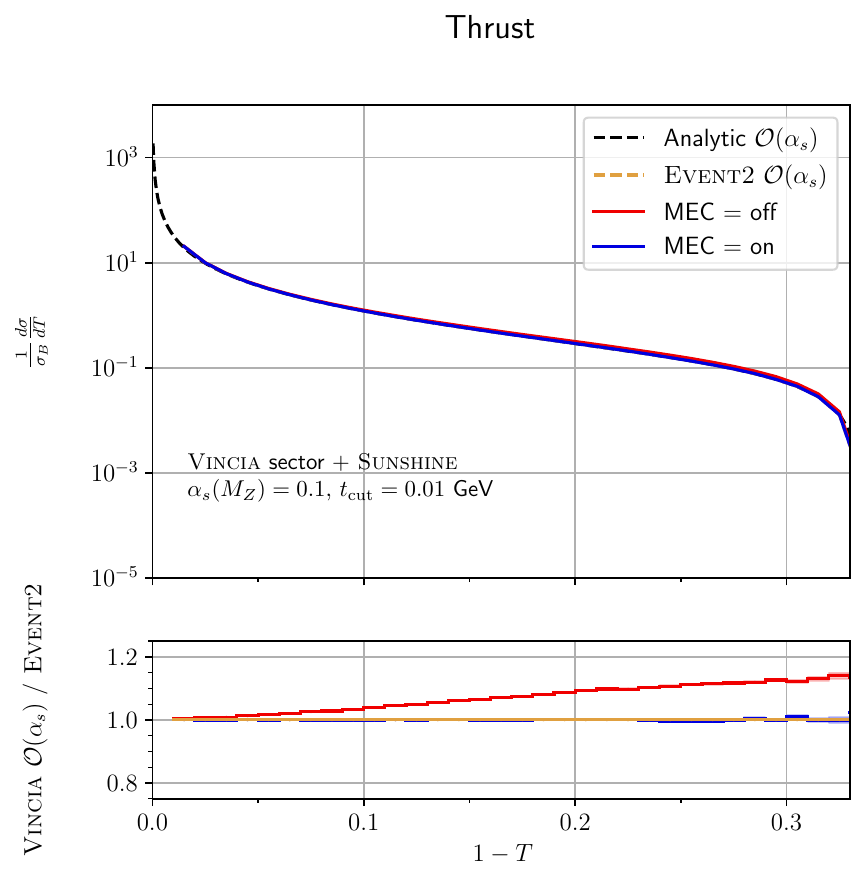}%
    \includegraphics*[width=0.499\textwidth,page=4]{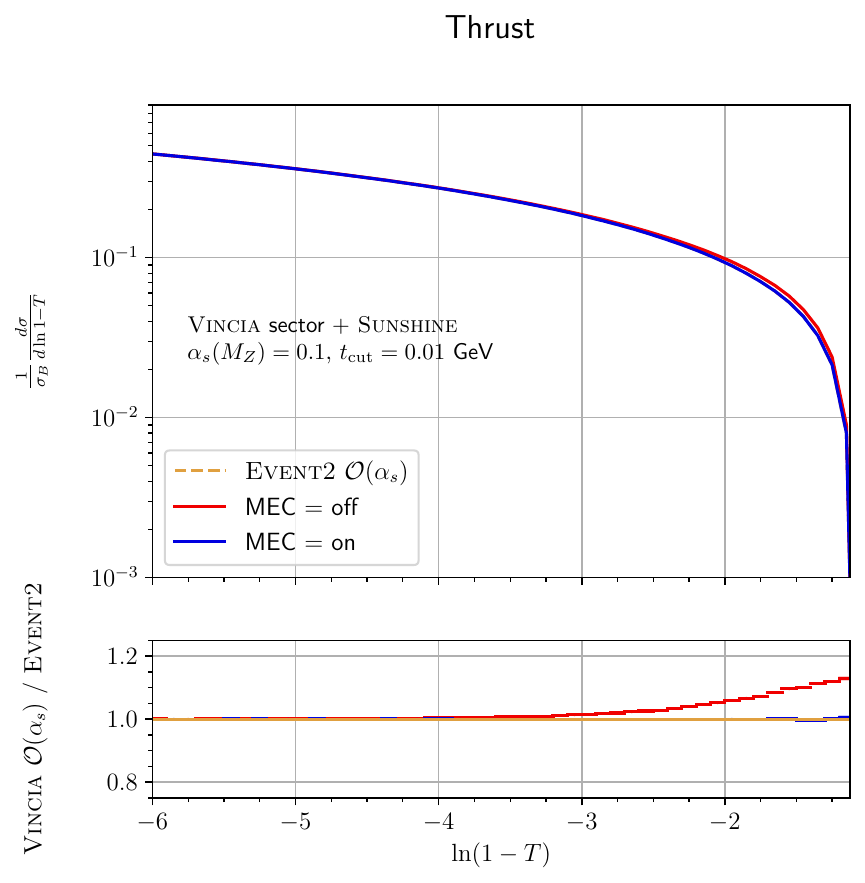}
    \caption{Tests of the $\mathcal{O}(\alpha_s)$ expansion of the \Vincia sector shower, for the thrust $1-T$ (left) and $\ln(1-T)$ (right), without (red) and with (blue) MECs.}
    \label{fig:fo-tests-as1-thrust}
\end{figure}
\begin{figure}[tp]
    \centering
    \!\includegraphics*[width=0.499\textwidth,page=5]{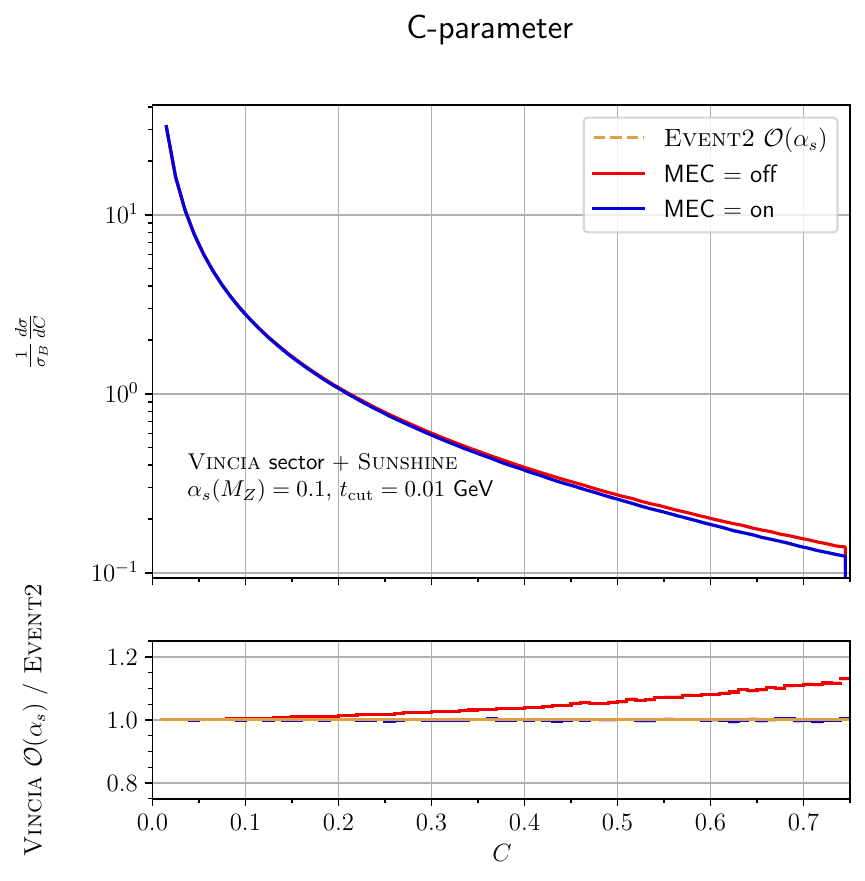}%
    \includegraphics*[width=0.499\textwidth,page=6]{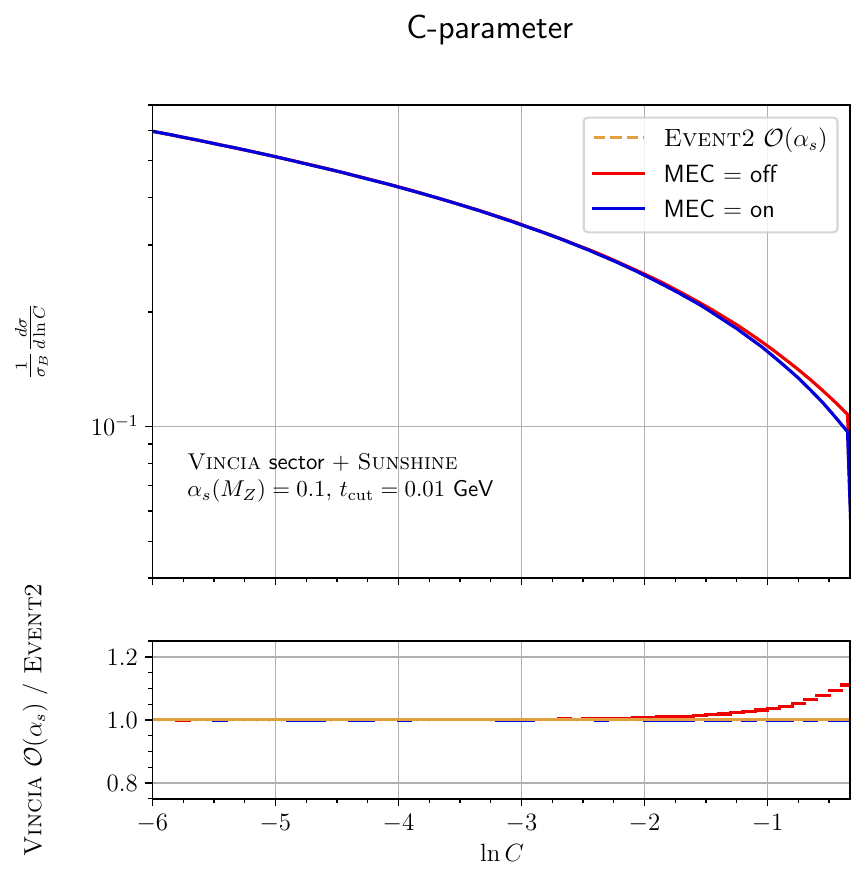}
    \caption{Same as fig.~\ref{fig:fo-tests-as1-thrust} for the $C$-parameter.}
    \label{fig:fo-tests-as1-C}
\end{figure}
Additionally, we show results for the distribution of the thrust~\cite{Brandt:1964sa,Farhi:1977sg} in fig.~\ref{fig:fo-tests-as1-thrust} and the $C$-parameter~\cite{Parisi:1978eg,Donoghue:1979vi,Ellis:1980wv} in fig.~\ref{fig:fo-tests-as1-C},
\begin{equation}
    1-T = 1-\max_{\vec{n}}\frac{\sum_i \vec{p_i}\cdot \vec{n}}{\sum_i \vec{p_i}}\,, \quad C = 3 \left( 1 - \frac{1}{2} \sum_{i,j}\frac{(p_i \cdot p_j)^2}{(p_i \cdot Q) (p_j \cdot Q)} \right)\,,
\end{equation}
where the sums run over all final-state particles and $Q$ is the total $4$-momentum in the event.
Both quantities receive contributions at non-zero values of the observable starting at next-to-leading order (i.e.~they are populated exclusively by three-parton events here, since we truncate the shower after one single emission).

Figs.~\ref{fig:fo-tests-as1-thrust} and~\ref{fig:fo-tests-as1-C} show results obtained by running \Vincia in \Sunshine, with MECs disabled (red) and enabled (blue), compared to analytic expressions~\cite{DeRujula:1978vmq} (dashed black), or numerical predictions from \EventTwo~\cite{Catani:1996jh,Catani:1996vz} (dashed gold) at NLO. Again, the discrepancy in the hard region ($1-T \to 1/3$, or $C \to 3/4$) is resolved once MECs are enabled. We also note that the MECs are very stable down to low values of the observables.

\subsection{Second-order tests}
\label{sec:fo-tests-as2}

We perform a similar series of tests one order higher, at the level of $\mathcal{O}(\alpha_s^2)$ relative to the Born event. This is non-trivial, as such a comparison effectively tests that (1) the MECs function correctly in a nested way, (2) the full four-parton phase space is covered, i.e.~the nested $2\mapsto 3$ branchings are properly sectorised and the (unordered) direct $2\mapsto 4$ branchings complement that phase space appropriately, and (3) the four-parton matrix elements have more structure than one order lower, i.e.~they involve separate channels ($q\bar q g g$, same-flavour $q\bar q q \bar q$ and mixed-flavour $q\bar q q' \bar q'$ final states), and subleading-colour contributions.

To perform these tests we consider observables that get contributions exclusively from double-real emission.
In particular we will first examine the $D$-parameter~\cite{Parisi:1978eg,Donoghue:1979vi,Ellis:1980wv}, which is defined as
\begin{equation}
D = 27 \det \Theta\,, \quad \Theta^{\alpha \beta} = \frac{1}{\sum_f |\vec p_f|} \sum_f \frac{p_f^\alpha p_f^\beta}{|\vec p_f|}\,,
\end{equation}
where the momentum tensor $\Theta$ is determined by summing over final-state particles $f$, with components $\alpha, \beta \in \lbrace 1,2,3\rbrace$ and $|\vec p_f|$ the modulus of their 3-momenta.

\begin{figure}[tp]
    \centering
    \!\includegraphics*[width=0.499\textwidth,page=1]{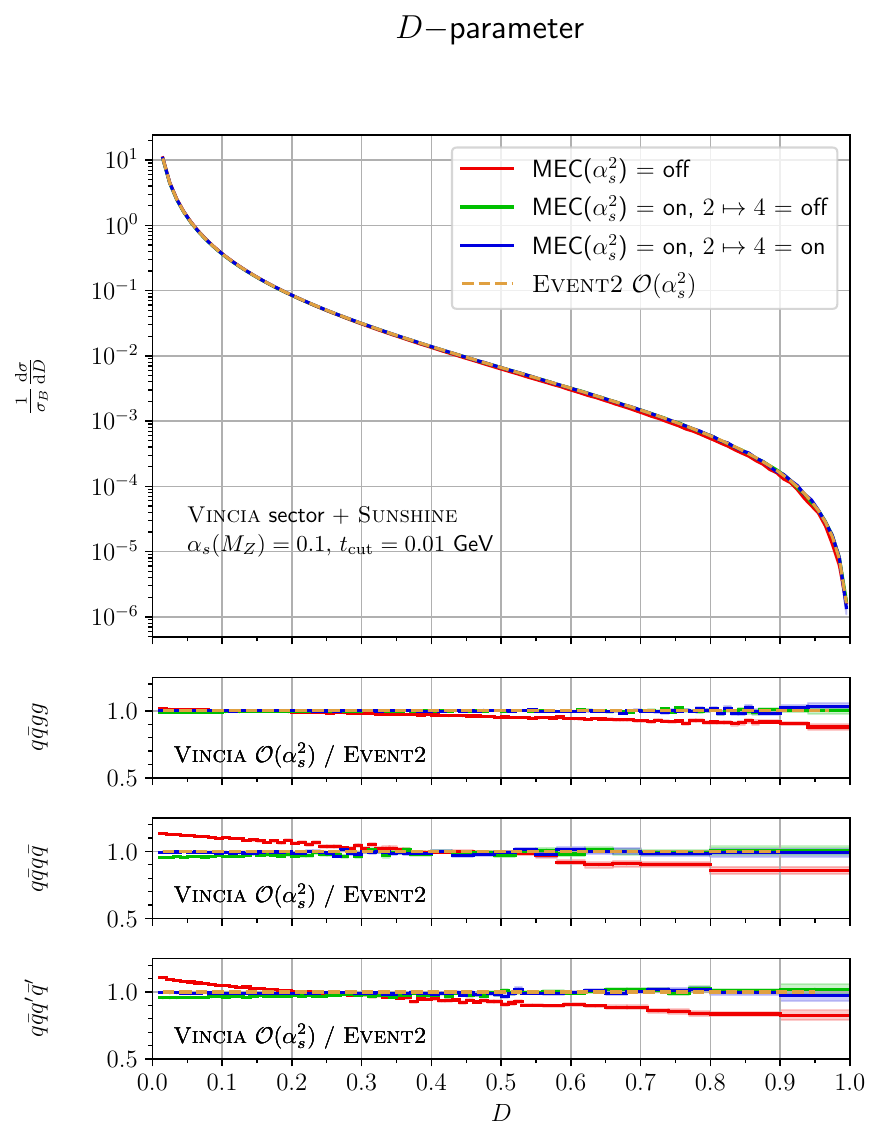}%
    \includegraphics*[width=0.499\textwidth,page=2]{figures/plot-test06-as2-v2.pdf}
    \caption{Tests of the $\mathcal{O}(\alpha_s^2)$ expansion for the $D$-parameter (left) and $\ln D$ (right). Results are shown without MECs for the second emission (red), with MECs but direct $2\mapsto 4$ branchings disabled (green), and with both enabled (blue). The contributing channels ($q \bar q g g$, $q\bar q q \bar q$ and $q\bar q q' \bar q'$) are separately compared to the \EventTwo results in the three ratio plots.}
    \label{fig:fo-tests-as2-D}
\end{figure}

In fig.~\ref{fig:fo-tests-as2-D} we compare the second-order expansion of the \Vincia shower, as obtained through \Sunshine, to the differential cross section from \EventTwo for the $D$-parameter. We show the total result in the main plot, and compare results to \EventTwo separately by flavour channel in the ratio plots, i.e.~for $q\bar qgg$, $q\bar q q' \bar q'$ with mixed-flavour quarks, $q \neq q'$, and for $q\bar q q \bar q$. The shower is run in three modes:
\begin{enumerate}
\item (red) with the first-order MEC, but leaving the second emission uncorrected,
\item (green) with both emissions corrected but without the direct $2\mapsto 4$ branchings enabled --- which implies that we should reproduce the correct matrix element for phase-space points reached by the parton shower, but the shower might miss some regions of phase space if they correspond to two unordered emissions, see fig.~\ref{fig:2to4},
\item (blue) with MECs up to and including two emissions, and (sectorised) nested $2\mapsto 3$ and direct $2\mapsto 4$ branchings.
\end{enumerate}
As we observe in fig.~\ref{fig:fo-tests-as2-D}, for $q\bar q gg$ (first ratio plot), the absence of the second-emission MEC (red) leads to a discrepancy of $-10\%$ in the hard region, $D\to 1$. At low scales, $D \ll 1$, the shower result also overshoots the exact prediction by about $\sim 2\%$, as evidenced in the logarithmic plot on the right of fig.~\ref{fig:fo-tests-as2-D}.\footnote{Note that this may be expected, as \Vincia does not reproduce the correct subleading-colour terms even at leading logarithmic accuracy (the same type of discrepancy pointed out in Ref.~\cite{Dasgupta:2018nvj} for $\alpha_s^2 L^4$ terms). Additionally at $\ln D \simeq -6$ the results are not completely dominated by leading logarithms.}
Without the required addition of the $2\mapsto 4$ branchings, the MEC-corrected shower (green curve) is in better agreement with the exact result, with no discernible difference in the hard region but a remaining discrepancy of about a percent in the $\ln D\to-\infty$ region. Only when both the two-emission MECs and the direct branchings are turned on, does the shower agree perfectly with the exact result.

Similar observations hold for four-quark final states (see the two last ratio plots), where the uncorrected curve (red) now differs from the correct result by a larger amount ($-20\%$ in the hard region, and $+15\%$ in the asymptotic region, $\ln D \to -\infty$).
The contribution of the direct $2\mapsto 4$ branchings is also larger there, about 5\% at small values of $D$, and brings the fully corrected result (blue) in agreement with \EventTwo.

We also examine observables that target distinct infrared regions of phase space, defined from Lund-plane~\cite{Andersson:1988gp,Dreyer:2018nbf} declusterings of the $4$-parton events. Specifically we focus on a region with two soft emissions of commensurate transverse momentum $k_{t,1} \simeq k_{t,2} \ll Q = M_Z$, and a region sensitive to spin correlations, both of which should not be described correctly by the pure shower approximation (since \Vincia implements neither the correct double-soft matrix elements nor spin correlations). In these regions, the shower corrected to second order is also expected to reproduce the exact results.

\begin{figure}[tp]
    \centering
    \raisebox{10pt}{\!\includegraphics[width=0.499\textwidth]{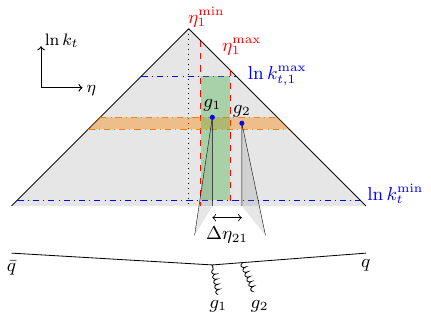}}%
    \includegraphics[width=0.499\textwidth]{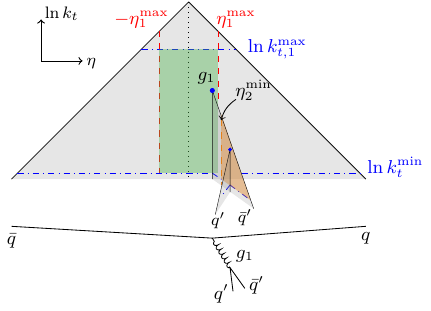}
    \caption{Illustration of the cuts performed to target two commensurately soft emissions on the primary Lund plane, for the $\Delta \eta_{21}$ observable (left); and a soft, large-angle primary splitting followed by a collinear secondary branching, for the $\Delta \psi_{12}$ observable (right). The shaded areas indicate the accepted phase spaces for the higher-$k_t$ (green) and lower-$k_t$ emissions (orange).
    }
    \label{fig:lund-plane-cuts}
\end{figure}

As a reminder, the Lund plane is constructed by clustering the event with the Cambridge algorithm~\cite{Dokshitzer:1997in} (which considers pairs of pseudojets that are closest in angle first), and recursively unravelling the clustering sequence back to the 2-jet event. Primary declusterings are obtained by following the harder branch at each iteration of the recursion. A secondary Lund leaf is defined by following the softer branch instead. Each declustering of two pseudojets, $i$ and $j$ (with $j$ the softer pseudojet, $|\vec{p}_i| > |\vec{p}_j|$), is associated with a value of transverse momentum $k_t$ and a rapidity $\eta$, defined as
\begin{equation}
k_t = |\vec{p}_j| \sin \theta_{ij}\,,\quad \eta = - \ln \tan \frac{\theta_{ij}}{2}\,,
\end{equation}
where $\theta_{ij}$ is the opening angle between the pseudojets.
We perform the Lund declustering of 4-parton events and define the following observables:

\begin{itemize}
    \item The difference in rapidity $\Delta \eta_{21} = \eta_2 - \eta_1$ between two primary Lund declusterings (``1'' and ``2''), where $1$ is the declustering with the highest $k_t$, i.e.~$k_{t,1} > k_{t,2}$.
    We require exactly two primary declusterings, and focus on a region where both declusterings are similarly soft (i.e.~double-soft emission), with the following set of cuts:
\begin{equation}
    -6 < \ln \frac{k_{t,1}}{Q} < -4\,,\quad 1 < \eta_1 < 3\,,\quad \ln \frac{k_{t,2}}{k_{t,1}} > -1\,.
\end{equation}
    \item The difference in azimuths $\Delta \psi_{12}$ as defined in Refs.~\cite{Karlberg:2021kwr,Hamilton:2021dyz}, sensitive to spin correlations, between the planes spanned by a soft, large-angle $q\to qg$ splitting on the primary Lund plane and a secondary, collinear splitting of the gluon (see e.g.~fig.~3 of Ref.~\cite{Karlberg:2021kwr}). We require exactly one primary (labelled ``1'') and one secondary (``2'') declustering, with
\begin{equation}
        -7 < \ln \frac{k_{t,1}}{Q} < -3\,,\quad |\eta_1| < 1\,,\qquad - 7 < \ln \frac{k_{t,2}}{Q}\,,\quad \eta_2 > 3\,.
\end{equation}
\end{itemize}
The definition of the observables is illustrated in fig.~\ref{fig:lund-plane-cuts}.
To evaluate these, we interfaced the \texttt{LundPlane}~\cite{Dreyer:2018nbf,Karlberg:2021kwr,Hamilton:2021dyz} code, v.~2.1.2, available from the FastJet~\cite{Cacciari:2011ma} contrib repository, to both \Vincia and \EventTwo. 

\begin{figure}[tp]
    \centering
    \!\includegraphics[width=0.499\textwidth,page=1]{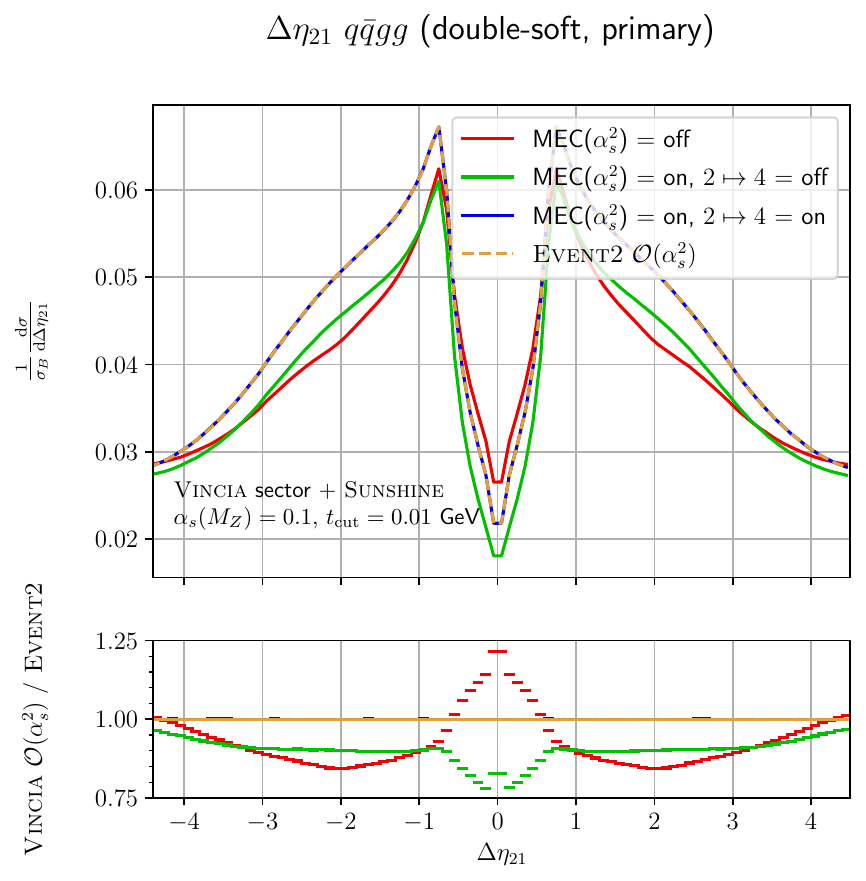}%
    \includegraphics[width=0.499\textwidth,page=5]{figures/plot-test06-as2.pdf}
    \caption{Same as fig.~\ref{fig:fo-tests-as2-D} for Lund declustering observables in the $q\bar q gg$ final state: (left) the difference in rapidity $\Delta \eta_{21}$ between two primary, soft declusterings, and (right) the azimuthal difference $\Delta \psi_{12}$ between a soft primary followed by a collinear secondary declustering. See text for the definition of the observables.
    }
    \label{fig:fo-tests-as2-y12-gg}
\end{figure}
Fig.~\ref{fig:fo-tests-as2-y12-gg} depicts $\Delta \eta_{21}$ in a double-soft region (left) and $\Delta \psi_{12}$ (right), for the $q\bar q g g$ final state. Firstly, in both cases we again observe that results with second-order MECs and direct $2\mapsto 4$ branchings agree perfectly with the exact results from \EventTwo. In the double-soft case (left-hand figure), the absence of $2\mapsto 4$ branchings (green curve) leads to a large discrepancy in the central region, of up to $\sim20\%$, which is completely resolved once the direct branchings are included (in blue).
On the right-hand side, we observe that the modulation of the distribution in $\Delta \psi_{12}$, which stems from spin correlations --- since $g\to gg$ splittings happen preferentially in-plane, i.e.~for $\Delta \psi = 0, \pm \pi$) --- is correctly reproduced once the second-order MECs are enabled.\footnote{In the single-logarithmic limit, $|\ln k_{t,1}/Q| \gg 1$ and $y_2 \gg 1$, we would expect the uncorrected curve (in red) to be completely flat, and the exact result to be of the form $\propto (1+B \cos (2\Delta \psi_{12}))$, with $B$ a positive constant. Numerical accuracy issues at very low scales make it difficult to push the \Vincia shower to similarly asymptotic regions as in Ref.~\cite{Hamilton:2021dyz}.}

\begin{figure}[tp]
    \centering
    \!\includegraphics[width=0.499\textwidth,page=2]{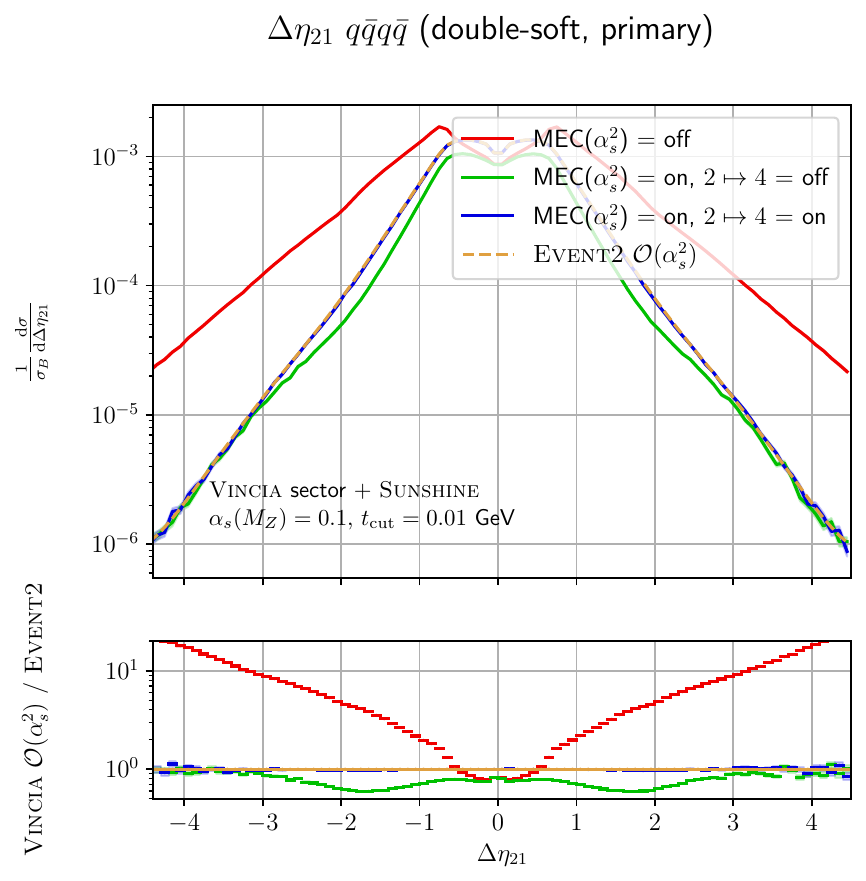}%
    \includegraphics[width=0.499\textwidth,page=8]{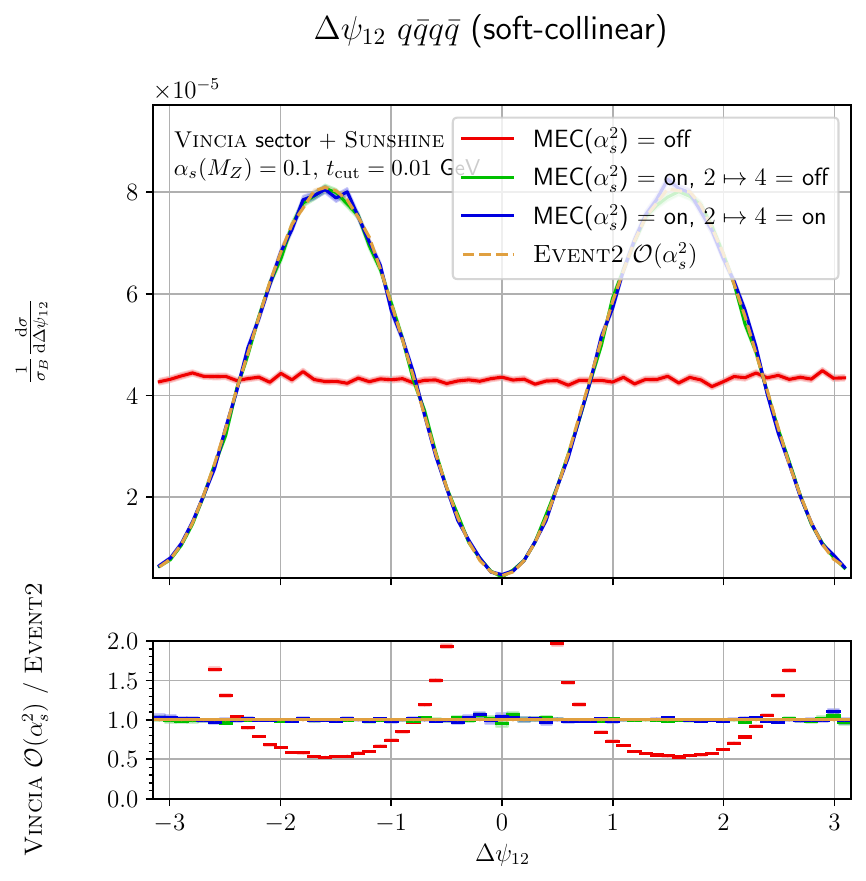}
    \caption{Same as fig.~\ref{fig:fo-tests-as2-y12-gg}, for the final state $q\bar q q\bar q$, with same-flavour quarks.
    }
    \label{fig:fo-tests-as2-y12-qq}
\end{figure}

\begin{figure}[tp]
    \centering
    \!\includegraphics[width=0.499\textwidth,page=3]{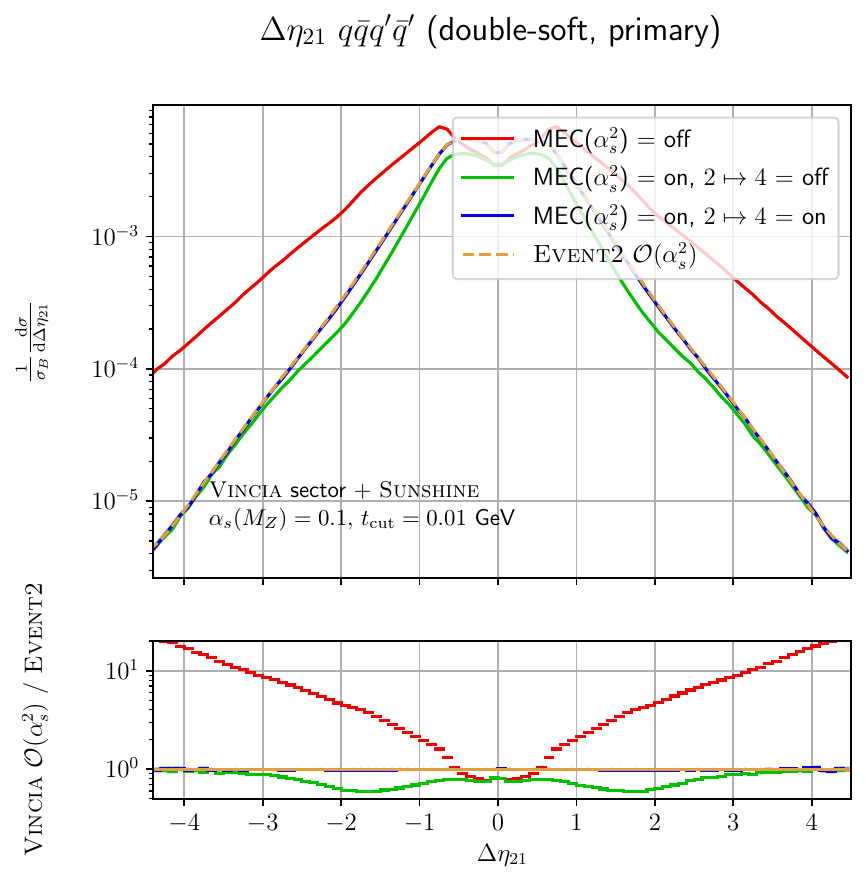}%
    \includegraphics[width=0.499\textwidth,page=11]{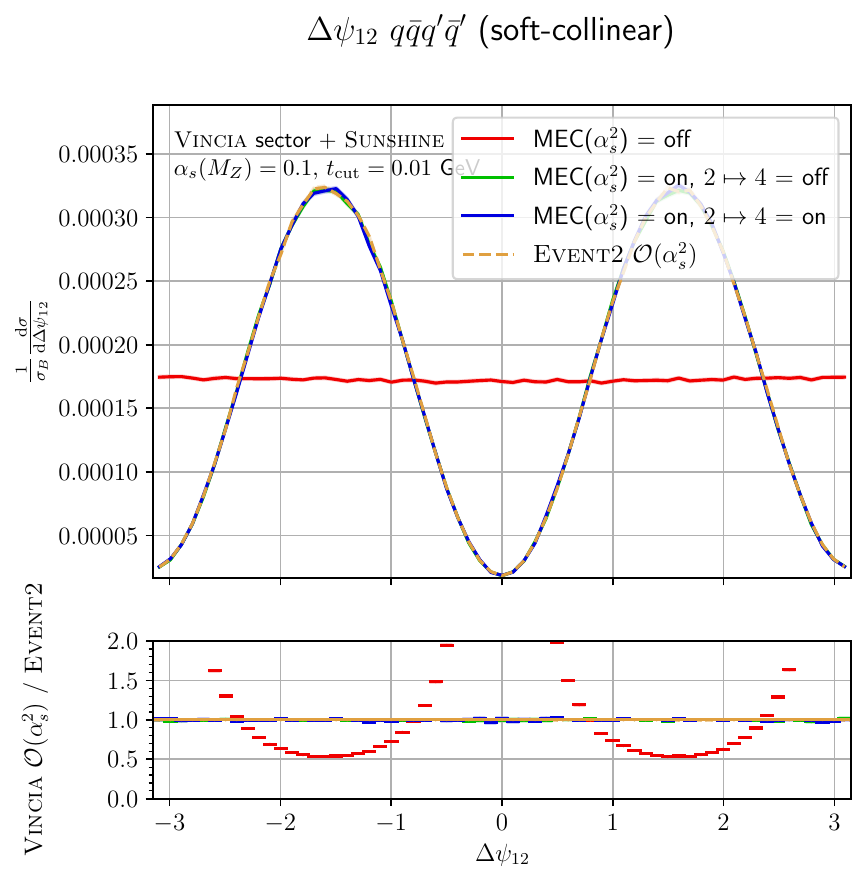}
    \caption{Same as fig.~\ref{fig:fo-tests-as2-y12-gg}, for the final state $q\bar q q'\bar q'$, with different-flavour quarks.
    }
    \label{fig:fo-tests-as2-y12-qqp}
\end{figure}

Similar conclusions can be drawn for the four-quark final states, see figs.~\ref{fig:fo-tests-as2-y12-qq} and~\ref{fig:fo-tests-as2-y12-qqp}. In the double-soft region (left), the impact of the matrix element corrections is much larger, as the scaling of the shower expansion at large rapidity differences, $|\Delta \eta_{21}| \gg 1$, is softer than that of the exact matrix element. Nevertheless, we again observe that the application of MECs at phase-space points reached by the shower is not enough, and one needs the contribution of the direct $2\mapsto 4$ branchings to restore agreement with the exact result.
In the case of spin correlations, as manifested in $\Delta \psi_{12}$ (right), the MEC curve also displays the expected behaviour --- this time the modulation peaks at $\Delta \psi = \pm \pi / 2$ as $g\to q\bar q$ splittings happen preferentially out-of-plane, and the strength of the modulation is much larger than for $g\to gg$ splittings, see e.g.~\cite{Richardson:2018pvo,Karlberg:2021kwr,Hamilton:2021dyz}.

In all comparisons presented above, the \Vincia sector shower, supplemented with MECs up to two emissions and direct $2\mapsto 4$ branchings, agrees with exact results when expanded with the help of \Sunshine. This constitutes an important test of the (tree-level aspects of the) NNLO MEC strategy proposed in~\cite{El-Menoufi:2024sys}.

\section{Summary and Outlook}\label{sec:conc}

In this article we have introduced a method, dubbed \Sunshine, to generate parton configurations distributed according to a parton shower's fixed-order expansion. We have implemented this algorithm in the framework of \textsc{Pythia}~8.3 and validated it in the context of \Vincia's sector shower. We then applied it to extract the sector-shower tree-level expansion and test the correctness of nested matrix-element corrections up to second order in $Z/\gamma^*$ hadronic decays. These tests critically rely on full coverage of phase space, which we achieve by (sectorised) nested $2\mapsto 3$ and direct $2\mapsto 4$ branchings.

We believe the proposed method and associated algorithm can be of broad use to enable easier and more direct studies of fixed-order expansions of shower algorithms, and to test the fidelity of strategies for matrix-element corrections, such as the framework for NNLO corrections proposed in~\cite{Campbell:2021svd,El-Menoufi:2024sys}. We also see a potential for constructive interplay between the non-unitary aspects introduced in \Sunshine and those of the recently proposed \textsc{Esme} strategy for NLO matching~\cite{vanBeekveld:2025lpz}. 

In addition, we intend to investigate whether the \textsc{Sunshine} algorithm could have further prospective uses, e.g., in the context of phase-space sampling for fixed-order calculations. So far, general-purpose shower algorithms have not been of direct use to fixed-order applications. Instead, proposals to exploit shower-like phase-space generation methods have been based on constructions of direct products of splitting kernels (and/or dipole/antenna functions), combined with shower-style $n\mapsto n+1$ kinematics maps. Such approaches have included \textsc{Sarge}~\cite{Draggiotis:2000gm,vanHameren:2000aj} and so-called forward-branching generators~\cite{Giele:2011tm,Figy:2018imt}. These have strong similarities with the algorithm proposed here, and we believe it would be well motivated to investigate to what extent a shower algorithm such as the \textsc{Vincia} sector shower run in the \textsc{Sunshine} mode could offer potentially competitive sampling methods for fixed-order applications more broadly. 

The \Sunshine code, including documentation and examples, will be publicly released in an upcoming version of \textsc{Pythia}~8.3. 

\subsection*{Acknowledgements}
We gratefully acknowledge stimulating discussions with Fabrizio Caola and Zara Rosenberg during the early stages of this work. We thank Gavin Salam and Jack Helliwell for comments on the manuscript, and Basem El-Menoufi for useful discussions and insight in the analytic four-quark matrix elements.
We also wish to thank the Aspen Center for Physics for hospitality and a wonderful working environment during the August 2024 ACP program on ``Tightening the Gap between Scattering Amplitudes and Events at the LHC at Higher Orders''. PS is supported by ARC grant DP220103512. LS is supported by an ARC Discovery Early Career Researcher Award (project number DE230100867). HTL is supported by the National Science Foundation of China under grant Nos.~12275156 and 12321005. The Aspen Center for Physics is supported by National Science Foundation Grant PHY-2210452.

\appendix

\section{Implementation of \Sunshine as a \texttt{UserHooks}}
\label{app:sunshine-algo}

In this appendix we give some details regarding the exact implementation of the \Sunshine algorithm within \textsc{Pythia}. While one could modify the shower algorithm itself to return events at each intermediate branching, we chose to implement \Sunshine using \textsc{Pythia}'s \texttt{UserHooks} structure. This allows complete flexibility, i.e.~in principle the \Sunshine hook can be used in conjunction with any \textsc{Pythia} shower module.

The main shower decision tree is represented in a simplified version in algorithm~\ref{alg:main-shower-loop} below. It is common to any parton shower (and for simplicity we omit details that are specific to \Vincia, e.g.~sector vetoes and MEC factors). The crucial point is that the \textsc{Pythia} framework allows the user to veto an emission, specifically through access to the \texttt{UserHooks:doVetoFSREmission} function, at the very end of the shower decision tree, as in line~\ref{alg:veto-call} of algorithm~\ref{alg:main-shower-loop}.

The emission veto is presented in algorithm~\ref{alg:veto-alg}, and the routine that provides the next \Sunshine event (e.g.~to be passed to an analysis) is contained in algorithm~\ref{alg:sunshine-next}.
In our implementation, we interface \Sunshine by \emph{always} vetoing the emission in the shower;\footnote{I.e.~we always return \texttt{true} in algorithm~\ref{alg:veto-alg}, cf.~also line~\ref{alg:emsn-veto-always-true} of algorithm~\ref{alg:main-shower-loop}.} we save the intermediate events, including the scale $t_n$ of the vetoed branching, into a buffer $\mathcal{B}$ of events (line~\ref{alg:save-intermediate} of algorithm~\ref{alg:veto-alg}); later, we treat the intermediate event by forcing the shower to generate a new trial emission, from the scale of the last branching $t_n$ down to the cutoff scale (line~\ref{alg:sunshine-force-emsn} of algorithm~\ref{alg:sunshine-next}).

We repeat that the pseudocode below is only one of many possible ways to implement \Sunshine in a given parton shower, and we adopted the \texttt{UserHooks} solution because it completely disentangles the \Sunshine algorithm from the main shower's.
Note that, starting from a given Born event, the main shower loop exits only when the scale of the next branching is below the shower cutoff, $t_{n+1} < t_\mathrm{cut}$, or when we reach the required number of emissions $n_\mathrm{max}$. The original Born event weight, $w_\mathrm{Born}$, is never modified by the procedure, i.e.~all bifurcations from the original event come with the same weight.

The \Sunshine code, algorithms~\ref{alg:veto-alg} and~\ref{alg:sunshine-next}, is contained in a header file which will be released in an upcoming version of \textsc{Pythia 8.3} together with an example file.

\begin{algorithm}
\caption{Main shower loop\label{alg:main-shower-loop}}
\begin{algorithmic}[1]
  \WHILE{$n_\mathrm{Born}^\mathrm{gen} < n_\mathrm{Born}^\mathrm{max}$}
    \STATE\label{alg:born-event} Generate a Born event $\mathcal{E}_\mathrm{Born}$ with the parton shower. Set $t_\mathrm{now} = t_0$ as the starting scale.\WHILE{$n_\mathrm{emsn} < n_\mathrm{max}$ or $t_\mathrm{now} < t_\mathrm{cut}$}
      \STATE\label{alg:trial-emsn} Generate a trial emission, with scale $t_n \in \left[ t_\mathrm{now},0 \right]$, with accept probability $p_\mathrm{accept}$.
      \IF{$p_\mathrm{accept} > \mathcal{R}_{[0,1]}$}
        \STATE Construct the $(n+1)$-parton event $\mathcal{E}$.
        \STATE\label{alg:veto-call} Call the \texttt{UserHooks} emission veto function, i.e.~go to~\ref{alg:veto-alg-1} of algorithm~\ref{alg:veto-alg}.
        \IF{\texttt{doVetoFSREmission} returns \texttt{true}}\label{alg:veto-emsn-true}
          \STATE Go back to \ref{alg:trial-emsn} to generate a new trial emission, with $t_\mathrm{now} = t_n$.
        \ELSE
        \STATE\label{alg:emsn-veto-always-true} // Do nothing. The emission veto always returns \texttt{true}.
        \ENDIF
      \ELSE
        \STATE Reject the trial emission and go back to~\ref{alg:trial-emsn}, with $t_\mathrm{now} = t_n$.
      \ENDIF
    \ENDWHILE
  \ENDWHILE
\end{algorithmic}
\end{algorithm}

\begin{algorithm}
\caption{{\tt SunshineHook}: Emission veto \texttt{doVetoFSREmission}\label{alg:veto-alg}}
\begin{algorithmic}[1]
  \STATE\label{alg:veto-alg-1} Read in the event $\mathcal{E}$ generated by the parton shower.
  \IF{$n_\mathrm{emsn} < n_\mathrm{max}$}
    \STATE\label{alg:save-intermediate} Save the event into the buffer $\mathcal{B}$ of intermediate events, and record the scale of the last emission, $\mathcal{B} = \mathcal{B} \cup \lbrace \mathcal{E}, t_n \rbrace$.
  \ELSE
  \STATE Save the event into the list $\mathcal{S}$ of \Sunshine events, $\mathcal{S} = \mathcal{S} \cup \lbrace \mathcal{E} \rbrace$.
  \ENDIF
\RETURN \texttt{true}, i.e.~go to~\ref{alg:veto-emsn-true} of algorithm~\ref{alg:main-shower-loop}.
\end{algorithmic}
\end{algorithm}

\begin{algorithm}
\caption{{\tt SunshineHook}: Event generation {\tt next()}\label{alg:sunshine-next}}
\begin{algorithmic}[1]
  \WHILE{\texttt{true}}
    \IF{$\mathcal{S}$ not empty}
      \STATE Pop the last event $\mathcal{E}$ from the list $\mathcal{S}$.
      \RETURN$\mathcal{E}$.
    \ELSIF{$\mathcal{B}$ not empty}
      \STATE\label{alg:sunshine-force-emsn} Pop the last event $\lbrace \mathcal{E}, t_n \rbrace$ from the buffer $\mathcal{B}$, and force one emission from the saved scale of the last branching, i.e.~go to~\ref{alg:trial-emsn} of algorithm~\ref{alg:main-shower-loop} with $t_\mathrm{now} = t_n$.
    \ELSIF{$n_\mathrm{Born}^\mathrm{gen} < n_\mathrm{Born}^\mathrm{max}$}
      \STATE Generate a new Born event $\mathcal{E}_\mathrm{Born}$ with the parton shower, i.e.~go to~\ref{alg:born-event} of algorithm~\ref{alg:main-shower-loop}.
    \ELSE
      \RETURN \texttt{false} // No more events to generate.
    \ENDIF
  \ENDWHILE
\end{algorithmic}
\end{algorithm}

\newpage

\bibliographystyle{SciPost_bibstyle}
\bibliography{main}

\begin{thebibliography}{10}
\providecommand{\url}[1]{\texttt{#1}}
\providecommand{\urlprefix}{}
\expandafter\ifx\csname urlstyle\endcsname\relax
  \providecommand{\doi}[1]{doi:\discretionary{}{}{}#1}\else
  \providecommand{\doi}{doi:\discretionary{}{}{}\begingroup
  \urlstyle{rm}\Url}\fi
\providecommand{\eprint}[2][]{\url{#2}}

\bibitem{Nagy:2009re}
Z.~Nagy and D.~E. Soper,
\newblock \emph{{Final state dipole showers and the DGLAP equation}},
\newblock JHEP \textbf{05}, 088 (2009),
\newblock \doi{10.1088/1126-6708/2009/05/088},
\newblock \eprint{0901.3587}.

\bibitem{Skands:2009tb}
P.~Z. Skands and S.~Weinzierl,
\newblock \emph{{Some remarks on dipole showers and the DGLAP equation}},
\newblock Phys. Rev. D \textbf{79}, 074021 (2009),
\newblock \doi{10.1103/PhysRevD.79.074021},
\newblock \eprint{0903.2150}.

\bibitem{Giele:2011cb}
W.~T. Giele, D.~A. Kosower and P.~Z. Skands,
\newblock \emph{{Higher-Order Corrections to Timelike Jets}},
\newblock Phys. Rev. D \textbf{84}, 054003 (2011),
\newblock \doi{10.1103/PhysRevD.84.054003},
\newblock \eprint{1102.2126}.

\bibitem{Dasgupta:2018nvj}
M.~Dasgupta, F.~A. Dreyer, K.~Hamilton, P.~F. Monni and G.~P. Salam,
\newblock \emph{{Logarithmic accuracy of parton showers: a fixed-order study}},
\newblock JHEP \textbf{09}, 033 (2018),
\newblock \doi{10.1007/JHEP09(2018)033},
\newblock [Erratum: JHEP 03, 083 (2020)],
\newblock \eprint{1805.09327}.

\bibitem{Forshaw:2020wrq}
J.~R. Forshaw, J.~Holguin and S.~Pl\"atzer,
\newblock \emph{{Building a consistent parton shower}},
\newblock JHEP \textbf{09}, 014 (2020),
\newblock \doi{10.1007/JHEP09(2020)014},
\newblock \eprint{2003.06400}.

\bibitem{Preuss:2024vyu}
C.~T. Preuss,
\newblock \emph{{A partitioned dipole-antenna shower with improved transverse
  recoil}},
\newblock JHEP \textbf{07}, 161 (2024),
\newblock \doi{10.1007/JHEP07(2024)161},
\newblock \eprint{2403.19452}.

\bibitem{Hoche:2024dee}
S.~H\"oche, F.~Krauss and D.~Reichelt,
\newblock \emph{{Alaric parton shower for hadron colliders}},
\newblock Phys. Rev. D \textbf{111}(9), 094032 (2025),
\newblock \doi{10.1103/PhysRevD.111.094032},
\newblock \eprint{2404.14360}.

\bibitem{vanBeekveld:2024wws}
M.~van Beekveld \emph{et~al.},
\newblock \emph{{New Standard for the Logarithmic Accuracy of Parton Showers}},
\newblock Phys. Rev. Lett. \textbf{134}(1), 011901 (2025),
\newblock \doi{10.1103/PhysRevLett.134.011901},
\newblock \eprint{2406.02661}.

\bibitem{Gribov:1972ri}
V.~N. Gribov and L.~N. Lipatov,
\newblock \emph{{Deep inelastic e p scattering in perturbation theory}},
\newblock Sov. J. Nucl. Phys. \textbf{15}, 438 (1972),
\newblock [Yad. Fiz.15,781(1972)].

\bibitem{Altarelli:1977zs}
G.~Altarelli and G.~Parisi,
\newblock \emph{{Asymptotic Freedom in Parton Language}},
\newblock Nucl. Phys. \textbf{B126}, 298 (1977),
\newblock \doi{10.1016/0550-3213(77)90384-4}.

\bibitem{Dokshitzer:1977sg}
Y.~L. Dokshitzer,
\newblock \emph{{Calculation of the Structure Functions for Deep Inelastic
  Scattering and e+ e- Annihilation by Perturbation Theory in Quantum
  Chromodynamics.}},
\newblock Sov. Phys. JETP \textbf{46}, 641 (1977),
\newblock [Zh. Eksp. Teor. Fiz.73,1216(1977)].

\bibitem{Gustafson:1987rq}
G.~Gustafson and U.~Pettersson,
\newblock \emph{{Dipole Formulation of QCD Cascades}},
\newblock Nucl. Phys. B \textbf{306}, 746 (1988),
\newblock \doi{10.1016/0550-3213(88)90441-5}.

\bibitem{Catani:1996jh}
S.~Catani and M.~H. Seymour,
\newblock \emph{{The Dipole formalism for the calculation of QCD jet
  cross-sections at next-to-leading order}},
\newblock Phys. Lett. \textbf{B378}, 287 (1996),
\newblock \doi{10.1016/0370-2693(96)00425-X},
\newblock \eprint{hep-ph/9602277}.

\bibitem{Kosower:1997zr}
D.~A. Kosower,
\newblock \emph{{Antenna factorization of gauge theory amplitudes}},
\newblock Phys. Rev. D \textbf{57}, 5410 (1998),
\newblock \doi{10.1103/PhysRevD.57.5410},
\newblock \eprint{hep-ph/9710213}.

\bibitem{Gehrmann-DeRidder:2005btv}
A.~Gehrmann-De~Ridder, T.~Gehrmann and E.~W.~N. Glover,
\newblock \emph{{Antenna subtraction at NNLO}},
\newblock JHEP \textbf{09}, 056 (2005),
\newblock \doi{10.1088/1126-6708/2005/09/056},
\newblock \eprint{hep-ph/0505111}.

\bibitem{Bierlich:2022pfr}
C.~Bierlich \emph{et~al.},
\newblock \emph{{A comprehensive guide to the physics and usage of PYTHIA
  8.3}},
\newblock SciPost Phys. Codeb. \textbf{2022}, 8 (2022),
\newblock \doi{10.21468/SciPostPhysCodeb.8},
\newblock \eprint{2203.11601}.

\bibitem{Brooks:2020upa}
H.~Brooks, C.~T. Preuss and P.~Skands,
\newblock \emph{{Sector Showers for Hadron Collisions}},
\newblock JHEP \textbf{07}, 032 (2020),
\newblock \doi{10.1007/JHEP07(2020)032},
\newblock \eprint{2003.00702}.

\bibitem{Ellis:1980wv}
R.~K. Ellis, D.~A. Ross and A.~E. Terrano,
\newblock \emph{{The Perturbative Calculation of Jet Structure in e+ e-
  Annihilation}},
\newblock Nucl. Phys. B \textbf{178}, 421 (1981),
\newblock \doi{10.1016/0550-3213(81)90165-6}.

\bibitem{Catani:1996vz}
S.~Catani and M.~H. Seymour,
\newblock \emph{{A General algorithm for calculating jet cross-sections in NLO
  QCD}},
\newblock Nucl. Phys. \textbf{B485}, 291 (1997),
\newblock \doi{10.1016/S0550-3213(96)00589-5, 10.1016/S0550-3213(98)81022-5},
\newblock [Erratum: Nucl. Phys.B510,503(1998)],
\newblock \eprint{hep-ph/9605323}.

\bibitem{Boos:2001cv}
E.~Boos \emph{et~al.},
\newblock \emph{{Generic User Process Interface for Event Generators}},
\newblock In \emph{{2nd Les Houches Workshop on Physics at TeV Colliders}}
  (2001), \eprint{hep-ph/0109068}.

\bibitem{Alwall:2006yp}
J.~Alwall \emph{et~al.},
\newblock \emph{{A Standard format for Les Houches event files}},
\newblock Comput. Phys. Commun. \textbf{176}, 300 (2007),
\newblock \doi{10.1016/j.cpc.2006.11.010},
\newblock \eprint{hep-ph/0609017}.

\bibitem{Larkoski:2013yi}
A.~J. Larkoski, J.~J. Lopez-Villarejo and P.~Skands,
\newblock \emph{{Helicity-Dependent Showers and Matching with VINCIA}},
\newblock Phys. Rev. D \textbf{87}(5), 054033 (2013),
\newblock \doi{10.1103/PhysRevD.87.054033},
\newblock \eprint{1301.0933}.

\bibitem{Fischer:2017htu}
N.~Fischer, A.~Lifson and P.~Skands,
\newblock \emph{{Helicity Antenna Showers for Hadron Colliders}},
\newblock Eur. Phys. J. C \textbf{77}(10), 719 (2017),
\newblock \doi{10.1140/epjc/s10052-017-5306-7},
\newblock \eprint{1708.01736}.

\bibitem{Kosower:2003bh}
D.~A. Kosower,
\newblock \emph{{Antenna factorization in strongly ordered limits}},
\newblock Phys. Rev. D \textbf{71}, 045016 (2005),
\newblock \doi{10.1103/PhysRevD.71.045016},
\newblock \eprint{hep-ph/0311272}.

\bibitem{Larkoski:2009ah}
A.~J. Larkoski and M.~E. Peskin,
\newblock \emph{{Spin-Dependent Antenna Splitting Functions}},
\newblock Phys. Rev. D \textbf{81}, 054010 (2010),
\newblock \doi{10.1103/PhysRevD.81.054010},
\newblock \eprint{0908.2450}.

\bibitem{Lopez-Villarejo:2011pwr}
J.~J. Lopez-Villarejo and P.~Z. Skands,
\newblock \emph{{Efficient Matrix-Element Matching with Sector Showers}},
\newblock JHEP \textbf{11}, 150 (2011),
\newblock \doi{10.1007/JHEP11(2011)150},
\newblock \eprint{1109.3608}.

\bibitem{Skands:2020lkd}
P.~Skands and R.~Verheyen,
\newblock \emph{{Multipole photon radiation in the Vincia parton shower}},
\newblock Phys. Lett. B \textbf{811}, 135878 (2020),
\newblock \doi{10.1016/j.physletb.2020.135878},
\newblock \eprint{2002.04939}.

\bibitem{Mrenna:2016sih}
S.~Mrenna and P.~Skands,
\newblock \emph{{Automated Parton-Shower Variations in Pythia 8}},
\newblock Phys. Rev. D \textbf{94}(7), 074005 (2016),
\newblock \doi{10.1103/PhysRevD.94.074005},
\newblock \eprint{1605.08352}.

\bibitem{Kleiss:2016esx}
R.~Kleiss and R.~Verheyen,
\newblock \emph{{Competing Sudakov Veto Algorithms}},
\newblock Eur. Phys. J. C \textbf{76}(7), 359 (2016),
\newblock \doi{10.1140/epjc/s10052-016-4231-5},
\newblock \eprint{1605.09246}.

\bibitem{Hartgring:2013jma}
L.~Hartgring, E.~Laenen and P.~Skands,
\newblock \emph{{Antenna Showers with One-Loop Matrix Elements}},
\newblock JHEP \textbf{10}, 127 (2013),
\newblock \doi{10.1007/JHEP10(2013)127},
\newblock \eprint{1303.4974}.

\bibitem{Li:2016yez}
H.~T. Li and P.~Skands,
\newblock \emph{{A framework for second-order parton showers}},
\newblock Phys. Lett. B \textbf{771}, 59 (2017),
\newblock \doi{10.1016/j.physletb.2017.05.011},
\newblock \eprint{1611.00013}.

\bibitem{Catani:1990rr}
S.~Catani, B.~R. Webber and G.~Marchesini,
\newblock \emph{{QCD coherent branching and semiinclusive processes at large
  x}},
\newblock Nucl. Phys. B \textbf{349}, 635 (1991),
\newblock \doi{10.1016/0550-3213(91)90390-J}.

\bibitem{vanBeekveld:2025lpz}
M.~van Beekveld, S.~Ferrario~Ravasio, J.~Helliwell, A.~Karlberg, G.~P. Salam,
  L.~Scyboz, A.~Soto-Ontoso, G.~Soyez and S.~Zanoli,
\newblock \emph{{Logarithmically-accurate and positive-definite NLO shower
  matching}}  (2025),
\newblock \eprint{2504.05377}.

\bibitem{Lavesson:2008ah}
N.~Lavesson and L.~L{\"o}nnblad,
\newblock \emph{{Extending CKKW-merging to One-Loop Matrix Elements}},
\newblock JHEP \textbf{12}, 070 (2008),
\newblock \doi{10.1088/1126-6708/2008/12/070},
\newblock \eprint{0811.2912}.

\bibitem{Lonnblad:2012ix}
L.~L\"onnblad and S.~Prestel,
\newblock \emph{{Merging Multi-leg NLO Matrix Elements with Parton Showers}},
\newblock JHEP \textbf{03}, 166 (2013),
\newblock \doi{10.1007/JHEP03(2013)166},
\newblock \eprint{1211.7278}.

\bibitem{Sjostrand:1987su}
T.~Sj{\"o}strand and M.~van Zijl,
\newblock \emph{{A Multiple Interaction Model for the Event Structure in Hadron
  Collisions}},
\newblock Phys. Rev. D \textbf{36}, 2019 (1987),
\newblock \doi{10.1103/PhysRevD.36.2019}.

\bibitem{Plehn:2005cq}
T.~Plehn, D.~Rainwater and P.~Z. Skands,
\newblock \emph{{Squark and gluino production with jets}},
\newblock Phys. Lett. B \textbf{645}, 217 (2007),
\newblock \doi{10.1016/j.physletb.2006.12.009},
\newblock \eprint{hep-ph/0510144}.

\bibitem{Corke:2010zj}
R.~Corke and T.~Sj{\"o}strand,
\newblock \emph{{Improved Parton Showers at Large Transverse Momenta}},
\newblock Eur. Phys. J. C \textbf{69}, 1 (2010),
\newblock \doi{10.1140/epjc/s10052-010-1409-0},
\newblock \eprint{1003.2384}.

\bibitem{Campbell:2021svd}
J.~M. Campbell, S.~H\"oche, H.~T. Li, C.~T. Preuss and P.~Skands,
\newblock \emph{{Towards NNLO+PS matching with sector showers}},
\newblock Phys. Lett. B \textbf{836}, 137614 (2023),
\newblock \doi{10.1016/j.physletb.2022.137614},
\newblock \eprint{2108.07133}.

\bibitem{El-Menoufi:2024sys}
B.~K. El-Menoufi, C.~T. Preuss, L.~Scyboz and P.~Skands,
\newblock \emph{{Matching Z $\to$ Hadrons at NNLO with Sector Showers}}
  (2024),
\newblock \eprint{2412.14242}.

\bibitem{Giele:2007di}
W.~T. Giele, D.~A. Kosower and P.~Z. Skands,
\newblock \emph{{A simple shower and matching algorithm}},
\newblock Phys. Rev. D \textbf{78}, 014026 (2008),
\newblock \doi{10.1103/PhysRevD.78.014026},
\newblock \eprint{0707.3652}.

\bibitem{Catani:2001cc}
S.~Catani, F.~Krauss, R.~Kuhn and B.~R. Webber,
\newblock \emph{{QCD matrix elements + parton showers}},
\newblock JHEP \textbf{11}, 063 (2001),
\newblock \doi{10.1088/1126-6708/2001/11/063},
\newblock \eprint{hep-ph/0109231}.

\bibitem{Lonnblad:2001iq}
L.~L{\"o}nnblad,
\newblock \emph{{Correcting the color dipole cascade model with fixed order
  matrix elements}},
\newblock JHEP \textbf{05}, 046 (2002),
\newblock \doi{10.1088/1126-6708/2002/05/046},
\newblock \eprint{hep-ph/0112284}.

\bibitem{Brooks:2020mab}
H.~Brooks and C.~T. Preuss,
\newblock \emph{{Efficient multi-jet merging with the Vincia sector shower}},
\newblock Comput. Phys. Commun. \textbf{264}, 107985 (2021),
\newblock \doi{10.1016/j.cpc.2021.107985},
\newblock \eprint{2008.09468}.

\bibitem{Bauer:2020jay}
C.~W. Bauer, N.~L. Rodd and B.~R. Webber,
\newblock \emph{{Dark matter spectra from the electroweak to the Planck
  scale}},
\newblock JHEP \textbf{06}, 121 (2021),
\newblock \doi{10.1007/JHEP06(2021)121},
\newblock \eprint{2007.15001}.

\bibitem{Brooks:2021kji}
H.~Brooks, P.~Skands and R.~Verheyen,
\newblock \emph{{Interleaved resonance decays and electroweak radiation in the
  Vincia parton shower}},
\newblock SciPost Phys. \textbf{12}(3), 101 (2022),
\newblock \doi{10.21468/SciPostPhys.12.3.101},
\newblock \eprint{2108.10786}.

\bibitem{Dasgupta:2020fwr}
M.~Dasgupta, F.~A. Dreyer, K.~Hamilton, P.~F. Monni, G.~P. Salam and G.~Soyez,
\newblock \emph{{Parton showers beyond leading logarithmic accuracy}},
\newblock Phys. Rev. Lett. \textbf{125}(5), 052002 (2020),
\newblock \doi{10.1103/PhysRevLett.125.052002},
\newblock \eprint{2002.11114}.

\bibitem{Hamilton:2020rcu}
K.~Hamilton, R.~Medves, G.~P. Salam, L.~Scyboz and G.~Soyez,
\newblock \emph{{Colour and logarithmic accuracy in final-state parton
  showers}},
\newblock JHEP \textbf{03}(041), 041 (2021),
\newblock \doi{10.1007/JHEP03(2021)041},
\newblock \eprint{2011.10054}.

\bibitem{Karlberg:2021kwr}
A.~Karlberg, G.~P. Salam, L.~Scyboz and R.~Verheyen,
\newblock \emph{{Spin correlations in final-state parton showers and jet
  observables}},
\newblock Eur. Phys. J. C \textbf{81}(8), 681 (2021),
\newblock \doi{10.1140/epjc/s10052-021-09378-0},
\newblock \eprint{2103.16526}.

\bibitem{Hamilton:2021dyz}
K.~Hamilton, A.~Karlberg, G.~P. Salam, L.~Scyboz and R.~Verheyen,
\newblock \emph{{Soft spin correlations in final-state parton showers}},
\newblock JHEP \textbf{03}, 193 (2022),
\newblock \doi{10.1007/JHEP03(2022)193},
\newblock \eprint{2111.01161}.

\bibitem{vanBeekveld:2022zhl}
M.~van Beekveld, S.~Ferrario~Ravasio, G.~P. Salam, A.~Soto-Ontoso, G.~Soyez and
  R.~Verheyen,
\newblock \emph{{PanScales parton showers for hadron collisions: formulation
  and fixed-order studies}},
\newblock JHEP \textbf{11}, 019 (2022),
\newblock \doi{10.1007/JHEP11(2022)019},
\newblock \eprint{2205.02237}.

\bibitem{vanBeekveld:2022ukn}
M.~van Beekveld, S.~Ferrario~Ravasio, K.~Hamilton, G.~P. Salam, A.~Soto-Ontoso,
  G.~Soyez and R.~Verheyen,
\newblock \emph{{PanScales showers for hadron collisions: all-order
  validation}},
\newblock JHEP \textbf{11}, 020 (2022),
\newblock \doi{10.1007/JHEP11(2022)020},
\newblock \eprint{2207.09467}.

\bibitem{Herren:2022jej}
F.~Herren, S.~H\"oche, F.~Krauss, D.~Reichelt and M.~Schoenherr,
\newblock \emph{{A new approach to color-coherent parton evolution}},
\newblock JHEP \textbf{10}, 091 (2023),
\newblock \doi{10.1007/JHEP10(2023)091},
\newblock \eprint{2208.06057}.

\bibitem{vanBeekveld:2023chs}
M.~van Beekveld and S.~Ferrario~Ravasio,
\newblock \emph{{Next-to-leading-logarithmic PanScales showers for Deep
  Inelastic Scattering and Vector Boson Fusion}},
\newblock JHEP \textbf{02}, 001 (2024),
\newblock \doi{10.1007/JHEP02(2024)001},
\newblock \eprint{2305.08645}.

\bibitem{MPFR}
L.~Fousse, G.~Hanrot, V.~Lef\`{e}vre, P.~P\'{e}lissier and P.~Zimmermann,
\newblock \emph{Mpfr: A multiple-precision binary floating-point library with
  correct rounding},
\newblock ACM Trans. Math. Softw. \textbf{33}(2), 13–es (2007),
\newblock \doi{10.1145/1236463.1236468}.

\bibitem{hida2000quad}
Y.~Hida, X.~S. Li and D.~H. Bailey,
\newblock \emph{Quad-double arithmetic: Algorithms, implementation, and
  application},
\newblock In \emph{15th IEEE Symposium on Computer Arithmetic}, pp. 155--162
  (2000).

\bibitem{vanBeekveld:2023ivn}
M.~van Beekveld \emph{et~al.},
\newblock \emph{{Introduction to the PanScales framework, version 0.1}},
\newblock SciPost Phys. Codeb. \textbf{2024}, 31 (2024),
\newblock \doi{10.21468/SciPostPhysCodeb.31},
\newblock \eprint{2312.13275}.

\bibitem{VinciaQCDSettings}
{VINCIA Collaboration},
\newblock \emph{{VINCIA QCD Antenna Shower Settings}},
\newblock \urlprefix\url{https://pythia.org/latest-manual/VinciaQCD.html},
  (accessed on 23-06-2025).

\bibitem{Brandt:1964sa}
S.~Brandt, C.~Peyrou, R.~Sosnowski and A.~Wroblewski,
\newblock \emph{{The Principal axis of jets. An Attempt to analyze high-energy
  collisions as two-body processes}},
\newblock Phys. Lett. \textbf{12}, 57 (1964),
\newblock \doi{10.1016/0031-9163(64)91176-X}.

\bibitem{Farhi:1977sg}
E.~Farhi,
\newblock \emph{{A QCD Test for Jets}},
\newblock Phys. Rev. Lett. \textbf{39}, 1587 (1977),
\newblock \doi{10.1103/PhysRevLett.39.1587}.

\bibitem{Parisi:1978eg}
G.~Parisi,
\newblock \emph{{Super Inclusive Cross-Sections}},
\newblock Phys. Lett. B \textbf{74}, 65 (1978),
\newblock \doi{10.1016/0370-2693(78)90061-8}.

\bibitem{Donoghue:1979vi}
J.~F. Donoghue, F.~E. Low and S.-Y. Pi,
\newblock \emph{{Tensor Analysis of Hadronic Jets in Quantum Chromodynamics}},
\newblock Phys. Rev. D \textbf{20}, 2759 (1979),
\newblock \doi{10.1103/PhysRevD.20.2759}.

\bibitem{DeRujula:1978vmq}
A.~De~Rujula, J.~R. Ellis, E.~G. Floratos and M.~K. Gaillard,
\newblock \emph{{QCD Predictions for Hadronic Final States in e+ e-
  Annihilation}},
\newblock Nucl. Phys. B \textbf{138}, 387 (1978),
\newblock \doi{10.1016/0550-3213(78)90388-7}.

\bibitem{Andersson:1988gp}
B.~Andersson, G.~Gustafson, L.~L{\"o}nnblad and U.~Pettersson,
\newblock \emph{{Coherence Effects in Deep Inelastic Scattering}},
\newblock Z. Phys. \textbf{C43}, 625 (1989),
\newblock \doi{10.1007/BF01550942}.

\bibitem{Dreyer:2018nbf}
F.~A. Dreyer, G.~P. Salam and G.~Soyez,
\newblock \emph{{The Lund Jet Plane}},
\newblock JHEP \textbf{12}, 064 (2018),
\newblock \doi{10.1007/JHEP12(2018)064},
\newblock \eprint{1807.04758}.

\bibitem{Dokshitzer:1997in}
Y.~L. Dokshitzer, G.~D. Leder, S.~Moretti and B.~R. Webber,
\newblock \emph{{Better jet clustering algorithms}},
\newblock JHEP \textbf{08}, 001 (1997),
\newblock \doi{10.1088/1126-6708/1997/08/001},
\newblock \eprint{hep-ph/9707323}.

\bibitem{Cacciari:2011ma}
M.~Cacciari, G.~P. Salam and G.~Soyez,
\newblock \emph{{FastJet User Manual}},
\newblock Eur. Phys. J. \textbf{C72}, 1896 (2012),
\newblock \doi{10.1140/epjc/s10052-012-1896-2},
\newblock \eprint{1111.6097}.

\bibitem{Richardson:2018pvo}
P.~Richardson and S.~Webster,
\newblock \emph{{Spin Correlations in Parton Shower Simulations}},
\newblock Eur. Phys. J. C \textbf{80}(2), 83 (2020),
\newblock \doi{10.1140/epjc/s10052-019-7429-5},
\newblock \eprint{1807.01955}.

\bibitem{Draggiotis:2000gm}
P.~D. Draggiotis, A.~van Hameren and R.~Kleiss,
\newblock \emph{{SARGE: An Algorithm for generating QCD antennas}},
\newblock Phys. Lett. B \textbf{483}, 124 (2000),
\newblock \doi{10.1016/S0370-2693(00)00532-3},
\newblock \eprint{hep-ph/0004047}.

\bibitem{vanHameren:2000aj}
A.~van Hameren and R.~Kleiss,
\newblock \emph{{Generating QCD antennas}},
\newblock Eur. Phys. J. C \textbf{17}, 611 (2000),
\newblock \doi{10.1007/s100520000508},
\newblock \eprint{hep-ph/0008068}.

\bibitem{Giele:2011tm}
W.~T. Giele, G.~C. Stavenga and J.-C. Winter,
\newblock \emph{{A Forward Branching Phase-Space Generator}}  (2011),
\newblock \eprint{1106.5045}.

\bibitem{Figy:2018imt}
T.~M. Figy and W.~T. Giele,
\newblock \emph{{A Forward Branching Phase Space Generator for Hadron
  colliders}},
\newblock JHEP \textbf{10}, 203 (2018),
\newblock \doi{10.1007/JHEP10(2018)203},
\newblock \eprint{1806.09678}.

\end{thebibliography}
\end{document}